\documentclass[useAMS,usenatbib]{mnras}

\usepackage[usenames,dvipsnames]{xcolor}

\usepackage{aas_macros}
\usepackage{amssymb,amsmath}
\usepackage{bm}
\usepackage{graphicx}
\usepackage{epstopdf}
\usepackage{aas_macros}
\usepackage{hyperref}
\hypersetup{colorlinks,citecolor=blue,linkcolor=blue,urlcolor=blue}

\title[Differential rotation and evolution of CFS-unstable stars]
{Long term evolution of CFS-unstable neutron stars and role of differential rotation on short time-scales
}

\author[A. I. Chugunov]
{A.~I.~Chugunov$^1$\thanks{andr.astro@mail.ioffe.ru}\\
    $^1$Ioffe Institute, St Petersburg, Russia
}

\begin{document}

%\begin{frontmatter}

\date{Accepted 2018 xxxx. Received 2018 xxxx;
    in original form 2018 xxxx}

\pagerange{\pageref{firstpage}--\pageref{lastpage}}
\pubyear{2018}

\maketitle

\label{firstpage}

%%%%%%%%%%%%%%%%%%%%%%%%%%%%%%%%%%%%%%%%%%%%%%%%%%%%%%%%%%%%%%%%%%%%%%
\begin{abstract}
I consider differential rotation, associated with radiation-driven Chandrasekhar-Friedman-Schutz (CFS) instability and its possible observational evidences. I focus on the evolution of the apparent spin frequency, which is typically associated with the motion of a specific point on the stellar surface (e.g., polar cap).
I start from long-term evolution (on the timescale when instability significantly changes the spin frequency). For this case, I reduce evolution equations to one differential equation and demonstrate that it can be directly derived from energy conservation law. This equation governs evolution rate through sequence of thermally equilibrium states and provides linear coupling for the cooling power and rotation energy losses via gravitational wave emission. In particular, it shows that differential rotation do not affect long-term spin-down. On the contrary, on the short timescales, differential rotation can significantly modify the apparent spin-down, if one examines a strongly unstable star with very small initial amplitude of unstable mode. This statement is confirmed by consideration of Newtonian non-magnetized perfect fluid and dissipative stellar models as well as magnetized stellar model.
For example, despite that widely applied evolution equations predict effective spin to be constant in absence of dissipation, the CFS-unstable star should be observed as spinning-down. However, effects of differential rotation on apparent spin-down are negligible for realistic models of neutron star recycling, unless: (1) the neutron star is not magnetized, and (2) r-mode amplitude is modulated faster than the shear viscosity dissipation timescale{, being} large enough that spin-down can be measured on modulation timescale.
\end{abstract}
%%%%%%%%%%%%%%%%%%%%%%%%%%%%%%%%%%%%%%%%%%%%%%%%%%%%%%%%%%%%%%%%%%%%%%

\begin{keywords}
    stars: neutron; instabilities; stars: evolution
\end{keywords}

\maketitle

\section{Introduction}

The  radiation-driven instability for rotating
stars  (Chandrasekhar-Friedman-Schutz, or CFS instability) was suggested by \cite{fs78a, fs78b}. It is essentially a perturbation enhancement in a rotating star by emission of gravitational waves (GWs).  For non-dissipative star models, the instability occurs at arbitrary rotation rate (\citealt{andersson98,fm98}). The dissipation suppresses instability up to a threshold frequency, which depends on temperature (e.g., \citealt*{lom98}). CFS instability is crucial for neutron stars (NSs) astrophysics, because for most rapidly rotating NSs it can take form of spontaneous excitation of r-modes. R-modes are specific type of oscillating modes, supported by the Coriolis force with predominantly toroidal type of oscillations. R-mode instability can alter  NSs evolution (\citealt*{levin99,hl00,btw07,as14,gck14a}), so that even a new class of neutron stars may emerge -- HOt and Fast Non-Accreting Rotators (HOFNARs) -- which stay hot for billions of years, thanks to r-mode instability (see \citealt*{cgk14} details). Neutron stars with unstable r-modes are considered to be a potential source for a GW astronomy (e.g., \citealt{ks16}). Even if real NSs are, in fact, stable throughout their lifetime, one can put observational constraints to the critical frequency of the instability and juxtapose them to the theoretical models of instability suppression. Thus, an opportunity to select more viable model would certainly provide new information on neutron stars (see, e.g., \citealt{haskell15} for recent review).

Up-to-date models of r-mode instability embrace full diversity of neutron star microphysics: they depend on NS core composition (e.g., \citealt*{Jones01_comment,lo02,no06,ahc10,as14_msp}), superfluidity (\citealt*{lm00,ac01,yl03a,yl03b};\ \ \  \citealt*{ly03,agh09,hap09,gck14b,gck14a,kg17,dkg18}), crust-core coupling (\citealt{rieutord01,lu01,ga06a,ga06b}),  in-medium effects (\citealt{kv15}), etc. To derive this microphysical information from observations properly, one should have an accurate theory, describing evolution of CFS unstable star observational features. This paper deals with one of elements of this theory -- role of differential rotation, which can be coupled to CFS instability.

The idea that GWs emission by CFS unstable star can generate differential
rotation was suggested by \cite{Spruit99_DifRot} within
a simple phenomenological model. Inevitable generation of differential rotation by r-mode instability in perfect fluid non-magnetized stellar models was confirmed by \cite*{lu01b,fll16}.
As discussed by \cite{hl00}, differential rotation complicates proper definition of the mean rotation rate $\hat \Omega$, which enters the evolution equations by \cite{olcsva98} and  \cite{hl00}.
In particular, in a recent paper (\citealt{Chugunov17}, henceforth Paper I) I pointed that
{
observations are typically sensitive to the secular motion of a specific point on the stellar surface (e.g., polar cap or hot spot),
% thus they are telltale for this motion,
corresponding to an apparent spin frequency $\Omega_\mathrm{app}$,
% (refered as observed frequency in Paper I)
which generally does not coincide with $\hat \Omega$ because of differential rotation.}
Furthermore, apparent spin down rate $\dot \Omega_\mathrm{app}$ can significantly differ from $\hat \Omega$ rate of change.

Hence, if one naively assumes them to be equal, the conclusions might be misleading.
Here I address a question of how one can correctly confront predictions of the r-mode evolution theory with
%{
the apparent spin down rate $\dot \Omega_\mathrm{app}$.
Note, that some recent papers suggest (e.g.,  \citealt*{kgc16}) or even make (e.g., \citealt{hp17,aaehh18}) such a comparison.

To answer this question,
I follow Paper I, where evolution equations were derived via multipolar expansion for GWs by \cite{Thorne80} (\citealt{olcsva98,hl00} appeal to the Lagrangian perturbation theory by \citealt{fs78a,fs78b}, while deriving evolution equations). This makes it possible to deal directly with Eulerian perturbations of the velocity. For completes, this derivation is presented (in a more detailed form) in Section \ref{Sec_GenEquations}.
In the Subsection \ref{Sec_longEvol} I demonstrate that on long timescales the evolution equations can be reduced to a one, which describes evolution rate through the sequence of thermally equilibrium states. Final Subsection \ref{Sec_DifRotGen} of Section \ref{Sec_EvolSimple} is devoted to the discussion, what is the effect of differential rotation on apparent frequency
{$\Omega_\mathrm{app}$, treated  as a variable which generally depends on the position of the observed point on NS surface. Real observation should reveal one of the possible values of this variable, corresponding to certain location of the observed point. I neglect other effects on the $\Omega_\mathrm{app}$, like the hot spot wandering discussed  \citealt{pw12}, assuming that they can be treated independently.}

I argue that the apparent spin down rate, averaged over long timescale, should not be affected by differential rotation. However, on a short timescale,  differential rotation contribution to the apparent spin down can be the same order of magnitude as the decrease rate of the mean rotation frequency.

Following Paper I, in Section \ref{Sec_slow} I illustrate the results considering r-mode instability in slow rotating Newtonian star models as an example and compare evolution equations with those widely applied (\citealt{olcsva98,hl00}). This section also provides a basis for a further discussion of differential rotation in Section \ref{Sec_DifRot}. As discussed in Paper I, at first glance, my evolution equations differ from those by \cite{olcsva98,hl00}, but this difference is shown to be spurious and associated with mean spin frequency definition. Namely, I define $\Omega$ as derivative of rotation energy over angular momentum, Eq.\ (\ref{Omega}),  but definition of $\hat \Omega$ by \cite{olcsva98,hl00} is corrected for canonical angular momentum of the mode. In particular, this divergence leads to different rate of change of $\Omega$ and $\hat \Omega$, which is the most striking in case of r-mode instability in non-dissipative stellar models: equations by \cite{olcsva98,hl00} predict $\hat \Omega=\mathrm{const}$, but $\Omega$ is decreasing, and the spin down rate is exponentially growing. Subsequent discussion in Section \ref{Sec_DifRot} reveals that the apparent frequency, in fact, decreases {for any position of the observed point}, even if there is no dissipation, thus favouring $\Omega$ and evolution equations of Paper I. I also discuss the term, suggested by \cite{hl00}, which couples evolution of mode amplitude with external (non r-mode) torques.

In Section \ref{Sec_DifRot} I discuss several models of differential rotation associated with r-modes as a testbench for general arguments from Subsection \ref{Sec_DifRotGen}. In particular, I apply the results by \cite{fll16}, obtained for unstable r-modes in perfect fluid Newtonian stellar models and demonstrate that the unstable star should be observed as spinning down at the rate $\dot \Omega_\mathrm{app}\sim (0.5\div 0.8)\dot \Omega<0$, dependent on the location of the observed point on the stellar surface (here and below numerical factors correspond to polytropic EOS, $P\propto \rho^{1+1/n}$ with $n=1$ and r-mode with azimuthal number $m=2$). In Subsection \ref{Sec_DifVisc} I examine differential rotation in dissipative non-magnetized star. Analysis is performed within generalized toy model by \cite{flrc17} and shows that differential rotation can affect apparent spin down rate for growing (unsaturated) and, in some cases, saturated r-modes (corresponding criteria are suggested). In Subsection \ref{Sec_DifMag} I discuss r-mode instability in magnetized perfect fluid star, applying the results by \cite{flrc17} (see \cite{cflr17} for short summary). They demonstrate that magnetic field suppresses differential motion of fluid elements. Thus, with a good reason I conclude that the apparent spin rate of change $\dot \Omega_\mathrm{app}$ should be equal to $2\dot \Omega/3<0$ for growing  modes and  to $\dot \Omega$ for saturated r-mode.

In the final Subsection \ref{Sec_DifRotRecyc} I qualitatively assess the role of differential rotation in a realistic scenario of a neutron star recycling by accretion in a low mass X-ray binaries (LMXBs). I argue that in this case the differential rotation can hardly affect the apparent spin down significantly because of the slow evolution of the r-mode amplitude.
The latter is associated with a specific feature of neutron star evolution in the instability window: the unstable mode is either saturated or star resides near the instability region's boundary. However, under specific (and rather unrealistic) conditions, the differential rotation can affect apparent spin down rate.
I conclude in Section \ref{Sec_DisConc}.

\section{CFS instability and the evolution of an unstable star}
\label{Sec_EvolSimple}

Here I discuss the evolution of CFS unstable relativistic rotating star in its asymptotic rest frame.
For completeness, in Subsection \ref{Sec_GenEquations} I present derivation of general evolution equations from paper I in more details. In Subsection \ref{Sec_longEvol} I discuss evolution on a long timescale. I show that the evolution equations are simplified and can be reduced to one differential equation.
The final Subsection  \ref{Sec_DifRotGen} is devoted to general discussion of differential rotation effects on the apparent spin frequency.

\subsection{General evolution equations in asymptotic rest frame}\label{Sec_GenEquations}
In asymptotic rest
frame total
mass-energy $E$ and angular momentum $J$ are well defined
(see, e.g., Section 19 by \citealt*{MTW}).
Uniform rotation corresponds to the minimal energy at a given angular momentum $J$  (e.g.\
\citealt{bl66,hs67,Stergioulas03})
and correspondent energy is rotational energy  $E_\mathrm{rot}$.
The rotation frequency can
be defined as
\begin{equation}
\Omega=\frac{\partial  E_\mathrm{rot}}{\partial J}.
\label{Omega}
\end{equation}

If total energy $E$ of the star exceeds $E_\mathrm{rot}$, the motion of fluid elements is perturbed with respect to uniform rotation.
Positively defined excitation energy $E_\mathrm{ex}$ can be attributed to the perturbation%
\footnote{$E_\mathrm{ex}$ should not be confused with
    canonical energy introduced by \cite{fs78a,fs78b}, see
    discussion in Section \ref{Sec_slow}.}
\begin{equation}
E_\mathrm{ex}=E-E_\mathrm{rot}(J)>0. \label{E_ex}
\end{equation}
Emission of the gravitational (or electromagnetic) waves
change  energy ($E\rightarrow E+\delta E$) and angular momentum ($J\rightarrow J+\delta J$) of the star (e.g., \citealt*{MTW}). Change of angular momentum causes variation of rotational energy to $E_\mathrm{rot}+\delta E_\mathrm{rot}$, where $\delta E_\mathrm{rot}=\Omega\delta J$.
There are three cases, if we compare $\delta E$ and $\delta E_\mathrm{rot}$:
\begin{description}
\item{$\delta E>\delta E_\mathrm{rot}$:} Excitation energy increases $\delta
E_\mathrm{ex}=\delta E-\delta E_\mathrm{rot}>0$. Even if a
star initially rotates uniformly, it does not in the final state:
perturbations should be excited. It is the mechanism of CFS
instability (see discussion below and Paper I);

\item{$\delta E=\delta E_\mathrm{rot}$:} variation of the excitation energy $\delta
E_\mathrm{ex}=0$. If initial star has uniform rotation
($E_\mathrm{ex}=0$), its final state is also uniform
rotation, but spin frequency changes. For example, it happens
during magneto-dipolar spin down of pulsar or GW emission by static mountains (see below);

\item{$\delta E<\delta E_\mathrm{rot}$:} Excitation energy decreases $\delta
E_\mathrm{ex}<0$. It occurs, only if uniform rotation was perturbed initially and a star's excitation energy is positive. Gravitational radiation acts as damping mechanism for perturbations. For example, it is the case for GW emission by oscillation modes in non-rotating neutron stars (e.g., \citealt{pt69,McDermott_etal84_OscEmEmis}).
\end{description}

Axisymmetric rotating star cannot emit GWs, so I consider further the emission associated with oscillation
modes. Indeed, if oscillation mode is excited with a frequency $\omega$ (in
the asymptotic rest frame) and azimuthal number $m$ [i.e. the
perturbations are $\propto \mathrm e^{\imath (\omega
t+m\phi)}$ and  cylindrical coordinate system
$(z,\,\varpi,\,\phi)$ with $\bm e_z=\bm J/J $ is applied], the emission
of GWs are permitted.

Multipolar expansion for GWs, formulated by \cite{Thorne80}, gives the rate of energy changes $\dot E^\mathrm{GW}$ and
angular momentum (in particular, for its projection to the rotation axis, $\dot
J^\mathrm{GW}$) due to GWs emission
as sums over multipolar contributions of radiation field in the local wave zone.
As it follows from Eqs.\ (4.16) and (4.23) by \cite{Thorne80}, $\dot E^\mathrm{GW}$ and $\dot
J^\mathrm{GW}$ are coupled by the equation, which is well-known for emission of electromagnetic
waves [see e.g.\ Section 9.8 by \cite{JacksonCED}]:
\begin{equation}
    -\frac{\omega}{m} \dot J^\mathrm{GW}=\dot
    E^\mathrm{GW}. \label{rate_GW_general}
\end{equation}
Here I discuss emission of GWs, thus the energy is carried away from the star: $\dot E^\mathrm{GW}<0$.
However, angular momentum of the star can either decrease or increase through emitted GWs, dependent on the sign of $\omega/m$. Definition of $E_\mathrm{ex}$ (Eq.\ \ref{E_ex}) permits me to write its rate of change as
\begin{equation}
  \dot E^\mathrm{GW}_\mathrm{ex}=\dot E^\mathrm{GW}-\dot E^\mathrm{GW}_\mathrm{rot}
  =\left(1+\frac{m\Omega}{\omega}\right) \dot E^\mathrm{GW}.
\label{Eex_GW}
\end{equation}
This means that excitation energy is increasing (and, consequently,
{GW emission enhances perturbations}%
) if and only if $(1+m\Omega/\omega)<0$. %
It is easy to check, that this condition is equal to the well-known criteria of CFS instability for eigenmodes: a mode is unstable if it is prograde in the inertial frame, but retrograde in a frame, corotating with the star (e.g. \citealt{ak01,fs11}).%
\footnote{The pattern speed of the mode is a speed of the surfaces of constant phase of $ e^{\imath (\omega t+m\phi)}$, given by $ \omega   t+m\phi=\mathrm{const}$. In the inertial frame it  leads to  $\sigma_\mathrm{p}^{\mathrm{in}}=\mathrm d\phi/\mathrm dt=-\omega/m$. For the  corotating frame, the azimuthal coordinate is $\phi^\mathrm{cor}=\phi-\Omega t$. The pattern speed in this frame is given by  $m\phi+\omega t=m\phi^\mathrm{cor}+(\omega+m\Omega)t=\mathrm{const}$ and can be written as $\mathrm d\phi^\mathrm{cor}/\mathrm d  t=\sigma_\mathrm{p}^{\mathrm{cor}}=(m\Omega+\omega)/m$.
Being combined, conditions $\sigma_\mathrm{p}^{\mathrm{in}}>0$ (prograde mode in inertial frame) and $\sigma_\mathrm{p}^{\mathrm{cor}}<0$ (retrograde mode in corotating frame) are equal to one condition $1+m\Omega/\omega<0$. Note, $\Omega>0$ due to selection of coordinate system.}
Note, if
asymmetry of the uniformly rotating star is static in the frame corotating with the star (e.g., it is associated with elasticity or magnetically supported mountains in crust), the associated GW emission does not enhance excitation
energy (at least while the star keeps its shape),
because emitted waves correspond to
$\omega=-m\Omega$ in asymptotic rest frame and thus $\dot E^\mathrm{GW}_\mathrm{ex}=0$ according to the equation (\ref{Eex_GW}). The same holds true for electromagnetic spin-down by magneto-dipolar emission (see, however, Section \ref{Sec_Compar} to  specify this statement for a case, which accounts for the effect of differential rotation).

To describe the evolution of CFS unstable star, I parametrize its state by three parameters: total angular momentum $J$,  mode energy $E_\mathrm{ex}$, and the parameter to characterize thermal state, which I choose to be temperature in the stellar centre $T$.
The latter implies an assumption that the star is thermally relaxed, which is applicable for NSs with high accuracy (e.g., \citealt*{plps04,gkyg05}) because of high thermal conductivity in their depths (see, e.g., \citealt*{sbs17,ss17} for recent results). Let me underline that I do not discuss thermal relaxation of the crust after accretion episodes, which was observed for many of transiently accreting NS (see \citealt*{wdp17} for recent observational review) and requires detailed modelling of temperature profile inside the crust (e.g., \citealt{Ruthledge_etal02_KS,syhp07,bc09,Ootes_etal16,Parikh_etal18_crustcool,mdkse18}).
{The latter effect is likely negligible for r-modes, because they are primarily localized in the core and temperature profile in the crust should not affect their properties considerably.}

The evolution equations, which uses the above parameters, are derived as follows. For angular momentum it follows from Eq.\ (\ref{Eex_GW}):
\begin{equation}
  \dot J^\mathrm{GW}
= -\frac{m}{\omega+m\Omega} \dot
  E^\mathrm{GW}_\mathrm{ex}. \label{J_evol_gen}
\end{equation}
In this
equation $\Omega=\Omega(J)$ and determined by Eq.\
(\ref{Omega}). It is worth to note that equation (\ref{J_evol_gen}) is applicable for
any oscillation mode
at any spin frequency and even for general relativistic
stellar models.

Evolution of the mode energy is associated with GW energy pumping  $\dot E^\mathrm{GW}_\mathrm{ex}$
and dissipation energy loses $\dot E^\mathrm{dis}_\mathrm{ex}$
\begin{equation}
\dot E_\mathrm{ex}=\dot E^\mathrm{GW}_\mathrm{ex}+\dot
     E^\mathrm{dis}_\mathrm{ex}.
     \label{Eex_evol_gen}
\end{equation}
Finally, the thermal evolution of star follows
\begin{equation}
  C \dot T=- \dot E^\mathrm{dis}_\mathrm{ex}
   -L_\mathrm{cool}.
 \label{thermal_gen}
\end{equation}
Here
$L_\mathrm{cool}$ and $C$ are
 total cooling power (neutrinos flux from bulk of the star and thermal emission from the
surface), and heat capacity of the star,
respectively (see, e.g., review by \cite{yp04} and \citealt{Ofengeim_etal17} for  useful analytical approximations).
If additional torque (e.g., accretion or magnetic braking) or heating processes (e.g., accretion- or rotation-induced deep crustal heating, \citealt{bbr98,gkr15}) affects to the star, corresponding torque and power $Q_\mathrm{other}$ should be added to the right parts of Eqs.\ (\ref{J_evol_gen}) and (\ref{thermal_gen}), respectively.

 To apply  equations (\ref{J_evol_gen})--(\ref{thermal_gen}), one should specify mode properties:
$\omega$, $\dot E^\mathrm{GW}_\mathrm{ex}$
and
 $\dot E^\mathrm{dis}_\mathrm{ex}$
as function of $E_\mathrm{ex}$, $J$ and $T$.%
\footnote{If dissipation is associated with the bulk
    viscosity, the perturbation energy can be directly converted to neutrino outflow (see, e.g.,
    \citealt*{rb03_bulk,gyg05, ams10}). This can be described by the above equations, but one should include enhancement of neutrino luminosity by perturbations to  $L_\mathrm{cool}$, so that it depend on excitation energy.}
Of course, this problem is very complicated, especially in the case of general relativity (see, e.g., \citealt{rk02,yl03b,lfa03,kgk10,jc17}).%

\subsection{Evolution on the long time-scale}\label{Sec_longEvol}
On the long timescales, or, to be more precise, on a timescales when
% the excitation and thermal energies can be neglected in comparison to {total} rotation energy change,
{excitation and thermal energies can be neglected in comparison to total change of the rotation energy $\Delta E_\mathrm{rot}$ over this timescale,
}
evolution equations can be simplified and reduced to one differential equation.

Indeed, on such timescales all energy pumped to the mode by instability is finally
converted to heat by dissipation (as a toy intuitive model, unstable mode works as a `tube'\
of negligible volume
to transfer
rotation energy to heat).
Thus, time-averaged heating power is equal to $\dot
E^\mathrm{GW}_\mathrm{ex}$  [one can prove it formally by
averaging of Eq.\ (\ref{Eex_evol_gen}) over time, while
neglecting the change of the mode energy between initial and final states]. Typically,
thermal evolution is much
faster than rotational evolution (because the rotation energy exceeds the thermal energy for many orders of magnitude, see, e.g., \citealt{as14}), and
the star, affected by CFS instability, rapidly evolves to the state of
thermal equilibrium, when the heating by CFS
instability and other mechanisms $Q_\mathrm{other}$ are
compensated by the cooling power $L_\mathrm{cool}$. In this
state Eq.\ (\ref{J_evol_gen}) can be rewritten as follows:
\begin{equation}
  \dot J^\mathrm{GW}=\frac{m}{\omega+m\Omega}\,\left(Q_\mathrm{other}-L_\mathrm{cool}\right).
   \label{Evol}
\end{equation}
Note, that the right part of this equation depends explicitly only on the $m$ and the mode frequency $\omega$, but not on  $E_\mathrm{ex}$, $\dot E^\mathrm{GW}_\mathrm{ex}$,
and
$\dot E^\mathrm{dis}_\mathrm{ex}$. The latter quantities, of course, enter this equation, but implicitly:
equation (\ref{Evol}) determines the evolution rate along
thermally equilibrium states on the
temperature-angular momentum (spin frequency) plane. The location of thermally equilibrium states on this plane
depends on $E_\mathrm{ex}$, $\dot E^\mathrm{GW}_\mathrm{ex}$, and
$\dot E^\mathrm{dis}_\mathrm{ex}$ and
 determined by a specific mechanism, which limits
CFS-instability. It can be nonlinear saturation (see e.g.
\citealt*{btw04a,btw04b,btw09,hga14}) or instability suppression through enhanced dissipation at specific temperature regions ($\left .\mathrm d \dot E^\mathrm{GW}_\mathrm{ex}/d T\right|_{E_\mathrm{ex}}>0$) (see, e.g., \citealt{ajk02,rb03,hap09,gck14a} for discussion of particular models). In the first
case, the nonlinear processes lead to rapid increase of $\dot E^\mathrm{dis}_\mathrm{ex}$ just as $E_\mathrm{ex}$ exceeds saturation energy $E^\mathrm{sat}_\mathrm{ex}$. It prevents the mode energy growing higher than $E^\mathrm{sat}_\mathrm{ex}$, which, in turn, determines the maximal value of
$\dot J^\mathrm{GW}$. The latter, via Eq.\
(\ref{Evol}), determines the cooling power (and thus, the temperature), which is required to keep the star in the thermal equilibrium.

In the second case, the star
moves along the boundary of the stability region (determined by  $\dot E^\mathrm{GW}_\mathrm{ex}+\dot E^\mathrm{dis}_\mathrm{ex}=0$). Location of this boundary determines stellar temperature and thus the cooling power. $\dot J^\mathrm{GW}$ (and thus the mode energy $E_\mathrm{ex}$) is adjusted  according to Eq.\
(\ref{Evol}) to provide enough heating to keep star at the required temperature (see, e.g., \citealt{gck14b,cgk14} for more detailed description).

Implicit dependence of  Eq.\ (\ref{Evol}) on $E_\mathrm{ex}$, $\dot E^\mathrm{GW}_\mathrm{ex}$
, and
$\dot E^\mathrm{dis}_\mathrm{ex}$ permits not only to model NS
evolution for certain theoretical models, which predict the sequence of thermal equilibrium states (as it was described above), but also to constrain the shape of the
instability windows and $\dot J^\mathrm{GW}$ from
observational data (e.g.\
\citealt{cgk17}).

It is worth noting, that
Eq.\ (\ref{Evol}) can be derived directly from conservation
laws formulated in Chapter 20 by \citealt{MTW}: `the rate of
loss of the 4-momentum and angular momentum from the
system, as measured gravitationally, is precisely equal to
the rate at which matter, fields and GWs
carry off 4-momentum and angular momentum'. The spatial
components of the stellar 4-momentum are vanishing, as far
as asymptotic rest frame is considered and cooling emission is
isotropic. Thus, the change of the total mass-energy is
associated with cooling (photon flux from  the surface and
neutrino flux from the bulk of the star) and GWs. If
thermal and excitation energies are negligible in
comparison to the change of rotational energy $\Delta E_\mathrm{rot}$, the change of
the total mass-energy can be estimated as the change of
rotational energy. Combined with Eq.\ (\ref{Eex_GW}), it
gives Eq.\ (\ref{Evol}), where power of additional energy
sources (e.g., accretion) $Q_\mathrm{other}$ is added artificially.

\subsection{Role of differential rotation} \label{Sec_DifRotGen}
As discussed by \citealt{Spruit99_DifRot,Rezzolla_etal00,lu01b,fll16}, the enhancement of CFS unstable mode can lead to differential rotation in the star.%
\footnote{
    \label{Foot_DifRot}
    Here by
    `differential rotation' I understand that oscillation-averaged Eulerian velocity differs from uniform rotation. As
    shown by \cite{Chugunov15}, that
    differential rotation does not necessarily leads to non-uniform secular motion of
    fluid elements in oscillating star, but angular
    velocity of this motion can differ from $\Omega$ (see
    discussion in Section \ref{Sec_DifMag}).}
{As discussed in introduction,}
the apparent spin
frequency $\Omega_\mathrm{app}$ is typically associated
with the motion of a certain point at the stellar surface (magnetic
poles, hot spots, etc.) and can differ from $\Omega(J)$ because of differential rotation.
To describe this effect, I
{treat $\Omega_\mathrm{app}$ as a variable, describing the spin frequency, which will be detected, if observation tracks a given point on the surface. Generally, it depends on the location of this point (see Sec.\ \ref{Sec_DifRot}). I introduce}
spin frequency correction for differential rotation
\begin{equation}
\Delta \Omega=\Omega_\mathrm{app}-\Omega.
\label{Om_obs}
\end{equation}
This correction is small ($\Delta \Omega\sim \Omega E_\mathrm{ex}/E_\mathrm{rot}\ll 1$, see e.g.\ \citealt{Rezzolla_etal00,lu01b,Sa04,Chugunov15,fll16}, obtained for Newtonian stellar models)%
\footnote{As discussed in Sections \ref{Sec_SaturNonmagId} and \ref{Sec_DifViscSatur}
for saturated r-modes in nonmagnetized Newtonian stellar models, differential rotation can exceed this estimate. See these sections for discussion of {respective} effects.} and clearly can be neglected if one is interested in the change of the spin frequency on a long timescale, as in Section \ref{Sec_longEvol}. However, on a short timescale the differential rotation may affect the apparent spin down rate. Namely,
\begin{equation}
\frac{\mathrm d \Omega_\mathrm{app}}{\mathrm d t}=
\frac{\mathrm d \Omega}{\mathrm d t}
+\frac{\mathrm d \Delta \Omega}{\mathrm d t}.
\label{dOm_obs}
\end{equation}
The first term in the right part is given by Eq.\ (\ref{J_evol_gen}) and can be estimated   as
\begin{equation}
\frac{\mathrm d \Omega}{\mathrm d t}\sim \Omega \frac{\dot J^\mathrm{GW}}{J}
    \sim \frac{\dot E_\mathrm{ex}^\mathrm{GW}}{J}.
\end{equation}
The second term{, by the order of magnitude, is}
%can be estimated as
%
\begin{equation}
\frac{\mathrm d \Delta \Omega}{\mathrm d t} \sim \Omega \frac{\dot E_\mathrm{ex}}{E_\mathrm{rot}}
\sim \frac{\dot E_\mathrm{ex}}{J}.
\end{equation}
For $\dot E_\mathrm{ex}\sim \dot E_\mathrm{ex}^\mathrm{GW}$, as it is suggested by Eq.\ (\ref{Eex_evol_gen}), both terms contributing to the apparent spin down
are of the same order of magnitude and thus {both can affect it}.
To study this effect accurately, one should not only specify the unstable mode and its properties, but also describe the evolution of differential rotation, which is the second-order effect in perturbation theory.  Within full general relativity this task is very complicated, and as for now, has not been performed yet self-consistently (see however \citealt{Kastaun11}, who discusses nonlinear decay of large amplitude r-modes to differential rotation).

In the following I limit consideration to r-mode instability in slowly rotating Newtonian perfect fluid stellar models, where several analytical results on the coupling of r-mode instability and differential rotation were recently obtained (\citealt{Sa04,Chugunov15,fll16,flrc17}).
In Section \ref{Sec_DifRot} these results and estimates for evolution of differential rotation in dissipative stellar models are applied as testbenches for the equations, suggested in this paper.

Differential rotation can also result in  evolution of the excitation energy through another torques, not associated with r-mode. Similar effect was suggested by \cite{hl00} within different approach (see discussion at the end of Section \ref{Sec_Compar}). As an example let me discuss a star affected by magnetic braking, estimated within the simplest magneto-dipolar model. As mentioned in Section \ref{Sec_GenEquations}, one can derive a counterpart of Eq.\ (\ref{Eex_GW}) for emission of electromagnetic waves. In case of magneto-dipolar emission, associated with rotation of the star, one should assume $m=-1$ and frequency of the emitted waves $\omega^\mathrm{M}$  to be equal to the magnetosphere spin velocity. In what follows I assume $\omega^\mathrm{M}=\Omega_\mathrm{app}$. According to Eq.\ (\ref{Eex_GW}) it gives
\begin{equation}
 \dot E_\mathrm{ex}^\mathrm{M}=\frac{\Delta \Omega}{\Omega}\dot E^\mathrm{M}, \label{dotE_mag}
\end{equation}
where $\dot E^\mathrm{M}=-E_\mathrm{rot}/\tau^\mathrm{M}$ and $\tau^\mathrm{M}>0$ are magnetic spin-down power and correspondent time-scale.

It should be noted, that this energy can be pumped not only to the unstable mode, but into differential rotation by itself.%
\footnote{Strictly speaking, Eq.\ (\ref{Eex_GW}) does not prove that the energy is pumped directly to the unstable mode, however, this assumption can be supported by general arguments: (a)  the pumping is resonant (at the frequency $\omega$ and multipolarity $m$ of the mode, which emits GW), (b) agreement with Lagrangian perturbation theory \cite{fs78a,fs78b,Friedman78}, and (c) analytic results by \cite{fll16}.}
Thus one had to be careful when including this term into evolution equations for unstable mode.
Note, $\dot E_\mathrm{ex}^\mathrm{M}$ depends on $\Delta \Omega$ and can be competitive with other terms in Eq.\ (\ref{Eex_evol_gen}), if  $\left|\dot E^\mathrm{M}\right|\gg\left|\dot E^\mathrm{GW}\right|$ or if differential rotation is strong enough $\Delta \Omega\sim \Omega$. In this paper I suppose that GW emission provides significant contribution to the spin down (e.g, $\left|\dot E^\mathrm{GW}\right|\gtrsim\left|\dot E^\mathrm{M}\right|$) and differential rotation is weak (because it is associated with second-order effect in the mode amplitude), thus, I neglect the terms associated with (\ref{dotE_mag}) below (except the discussion at the end of the next section).

\section{Evolution of r-mode unstable NS within slow rotating Newtonian stellar model}
\label{Sec_slow}

To illustrate the previous section and to provide basics for subsequent discussion of differential rotation, here I limit my
consideration to slow rotating Newtonian stellar models,
which are widely applied for studying CFS
instability in NSs (see, e.g., \citealt{haskell15} for
recent review). This analysis is close to the one presented in Paper I, but is slightly more detailed.
The effects of differential rotation on the apparent frequency are considered in Section \ref{Sec_DifRot}.
For slowly rotating case, Eq.\ (\ref{Omega}) can be written as $\Omega=J/I$, where moment of inertia $I$ does not depend on $J$. The most unstable mode is
r-mode with $l=m=2$ (e.g., \citealt{lom98}), which has
frequency
\begin{equation}
 \omega=-\frac{(m-1)(m+2)}{m+1}\Omega=-\frac{4}{3}\Omega.
 \label{om_r}
\end{equation}
For barotropic equation of state (pressure depends only on
density), the  first order Eulerian perturbations of
velocity are purely toroidal and can be written in the following form (e.g., \citealt{pp78})
\begin{equation}
 \delta^{(1)} \bm v=
 \sqrt{2}\alpha R\Omega \left(\frac{r}{R}\right)^m \mathfrak{Im} \bm Y_{mm}^B
 \exp^{i\omega t}
       \label{delta_v}
\end{equation}
Here  $\alpha$ is dimensionless mode amplitude, $\mathfrak{Im}(Z)$ is imaginary part of complex number $Z$, and
\begin{equation}
\bm Y_{lm}^B=\frac{1}{l(l+1)} r
  \nabla \times (r\nabla Y_{mm})
\end{equation}
is magnetic-type vector spherical harmonic (see e.g., \citealt*{vms88}).
Note, that
\cite{lom98,olcsva98} define $\delta^{(1)} \bm v$ as a complex value
and, as it was noted by \cite{lfa03}, calculate
energy
as a sum of respective values for the
real and imaginary parts. In Eq.\ (\ref{delta_v}) I define
$\delta^{(1)} \bm v$ as a real value and insert additional
multiplier $\sqrt 2$ to get the same normalization for the energy
in equation (\ref{E_ex_alpha}) as it was suggested by \cite{lom98,olcsva98} and  widely applied in subsequent papers on r-mode instability.
The Eulerian variations of density, pressure, and
gravitational potential are of the second order in
$\Omega$, and thus they  can be neglected in the leading order of slow rotation
approximation (see \citealt*{lmo99} for r-modes in the second order of slow rotation
approximation). As a result, only kinetic energy contributes to the excitation energy of r-mode, which can be written as
\begin{eqnarray}
    E_\mathrm{ex}&=&\int \frac{\rho \delta  v^2}{2} \mathrm d ^3
    \bm r
    =\int \frac{\rho [\delta^{(1)} \bm v]^2}{2} \mathrm d ^3
    \bm r
    \nonumber \\
    &+& \int \rho  \bm v_0 \delta^{(2)} \bm v  \mathrm d ^3
    \bm r
    +\mathcal O(\alpha^3). \label{E_slow_gen}
\end{eqnarray}
Here $\delta v^2=\delta( v^j v_j)$ and  $\delta \bm v=\sum_i\delta ^{(i)} \bm v$ are total
perturbation of velocity, presented as a sum over orders in
$\alpha$ [i.e., $\delta ^{(i)} \bm v=\mathcal O(\alpha^i)$]. Integral is taken over stellar volume. The second term in Eq.\
(\ref{E_slow_gen}) depends on the
second order velocity perturbation $\delta^{(2)}\bm v$, but only axisymmetric part   $\delta^{(2)}_\mathrm{sym}\bm v$ can contribute to the integral (due to axisymmetry of $\bm v^0$).
The axisymmetric part $\delta^{(2)}_\mathrm{sym}\bm v$, is
determined up to arbitrary cylindrically stratified differential rotation
(e.g.,\ \citealt{Sa04}).%
\footnote{To be more specific, as shown by  \cite{fll16}  for unstable r-modes in perfect fluid stellar models,  the  exponentially growing part of $\delta^{(2)}_\mathrm{sym}\bm v$ is unique, but it does not exclude contribution of time-independent differential rotation to $\delta^{(2)}_\mathrm{sym}\bm v$.}
However, definition of excitation energy adopted in this paper (Eq.\ \ref{E_ex}) requires
that given (perturbed) state of the star should be considered as perturbation of uniformly rotating star with same angular momentum. It imposes constraint:
\begin{equation}
    \delta J=\int \rho \left[\delta \bm v\times \bm
    r\right]\mathrm d^3 \bm r=0. \label{J_constraint}
\end{equation}
So far as unperturbed state is uniform rotation
$\bm v_0=\bm \Omega\times \bm r$,
$\delta^{(2)}_\mathrm{sym} \bm v$ contribution to the energy vanishes (in second
order in $\alpha$). Thus, the second order excitation energy
is determined exclusively by the first order perturbations and equals to
the kinetic energy in the system corotating with star, which is given by (see,
e.g., \citealt{lom98}):
\begin{equation}
    E_\mathrm{ex}=\int \frac{\rho [\delta^{(1)} \bm v]^2}{2} \mathrm d ^3
    \bm r=\frac{1}{2} \alpha^2 \Omega^2 R^{-2m+2} \int _0^R \rho r^{2m+2}\mathrm d^3
    \bm r.
    \label{E_ex_alpha}
\end{equation}

The instability timescale
\begin{equation}
    \tau^\mathrm{GW}=-2 \frac{E_\mathrm{ex}}{\dot E^\mathrm{GW}_\mathrm{ex}}
\end{equation}
can be calculated via multipolar expansion of gravitational
radiation for Newtonian sources (see Section V.C in
\citealt{Thorne80}), as it was done by \cite{lom98}.
Namely,
\begin{equation}
    \dot J^\mathrm{GW}=
    -m\omega
    \sum_{l=2}^{\infty}
    %\sum_m
    N_l\omega^{2l} \left(
    \left|\delta D_{lm}\right|^2+\left|\delta J_{lm}\right|^2       \right),
\end{equation}
where perturbations of the mass $\delta D^{lm}$ and current $\delta J^{lm}$ multipole moments are
\begin{eqnarray}
    \delta D^{lm}&=&
    \int \delta \rho\, Y_{lm}^\ast r^l \mathrm d^3 \bm r,
\\
\delta  J^{lm}&=&
    \frac{2}{c}\sqrt{\frac{l}{l+1}}
    \int  (\rho \delta \bm v+\delta \rho \bm v) \cdot \bm Y_{lm}^{B\ast} r^l \mathrm d^3 \bm r
\end{eqnarray}
Here
\begin{equation}
    N_l=
    \frac{4\pi G}{c^{2l+1}}
    \frac{(l+1)(l+2)}{l(l-1)[(2l+1)!!]^2}.
\end{equation}
After r-mode solution (Eq.\ \ref{delta_v}) is substituted into these equations, one obtains that dominant contribution to the instability timescale comes from current multiple (see e.g., \citealt{lom98,olcsva98}) and
\begin{eqnarray}
\frac{1}{\tau^\mathrm{GW}}
&=&-\frac{32\pi G\Omega^{2m+2}}{c^{2m+3}}
\,\frac{(m-1)^{2m}}{[(2m+1)!!]^2}
\,\left(\frac{m+2}{m+1}\right)^{2l+2}
 \nonumber \\
&\times& \int_0^R \rho r^{2m+2} \mathrm d r
\label{tau_GW}
\\
&\approx& -\frac{1}{10^3 \mathrm{\ s}}\, \left(\frac{R}{10\mathrm{\ km}}\right)^{4}
\left(\frac{M}{1.4M_\odot}\right)^{1}
\left(\frac{\nu}{600\mathrm{\ Hz}}\right)^{6}.
 \nonumber
\end{eqnarray}
Here $\nu=\Omega/2\pi$.
\cite{fll16} confirms this result by analytic
treatment of the r-mode instability up to second order in
oscillation amplitude for perfect fluid stellar model. In particular, this paper explicitly reveals the mechanism of energy pumping into the r-modes by GW back-reaction force.

 The dissipation rate
\begin{equation}
 \tau^\mathrm{dis}=-2\frac{E_\mathrm{ex}}{\dot E^\mathrm{dis}_\mathrm{ex}}
\end{equation}
can be calculated for certain model of dissipation as a sum over contributions of
relevant
dissipation processes
(shear viscosity, mutual friction, etc.). Note, any of internal dissipation processes cannot affect
total angular momentum of the star, thus the rotational energy is conserved. As a result $\dot E^\mathrm{dis}_\mathrm{ex}$ equals to dissipation rate of the total energy.
For example, the contribution of the shear viscosity $\eta$ to the dissipation rate is (\citealt{lom98})
\begin{eqnarray}
\frac{1}{\tau^\mathrm{S}}&=&(m-1)(2m+1)\frac{\int_0^R \eta r^{2l} \mathrm d r}
{\int_0^R \rho r^{2l+2} \mathrm d r}  \label{tau_S}
\\
&\approx& \frac{1}{2.2\times 10^5 \mathrm{\ s}} \,\left(\frac{R}{10\mathrm{\ km}}\right)^{-5}
     \left(\frac{M}{1.4M_\odot}\right)
      \left(\frac{T}{10^8\mathrm{\ K}}\right)^{-2}. \nonumber
\end{eqnarray}
The numerical factor corresponds to electron shear viscosity from \cite{sy08}, fitted by \cite{gck14a} for \cite{apr98} equation of state parametrized by \cite{hh99}. Critical temperature of proton superfluidity assumed to be $T_\mathrm{cp}=2\times10^9$\,K. As noted by \cite{gck14a}, coincidence of the viscosity with results by \cite{fi79} is accidental, because physics input is essentially different.

Introduction of above timescales permits me to rewrite Eqs.\ (\ref{J_evol_gen})
-- (\ref{thermal_gen}) in the form ($m=2$ is assumed, see Paper I):
\begin{eqnarray}
  \dot \Omega &=&\frac{2\tilde Q \alpha^2}{ \tau^\mathrm{GW}(\Omega)}\Omega \label{dot_Om} \\
  \dot \alpha &=& -\left(\frac{1}{\tau^\mathrm{GW}}+\frac{1}{\tau^\mathrm{dis}}\right) \alpha \label{dot_alpha} \\
  C\dot T &=& \frac{\tilde J M R^2 \Omega}{\tau^\mathrm{dis}} \alpha^2+
   Q_\mathrm{other}-L_\mathrm{cool} \label{dot_T}
\end{eqnarray}
Here I, following \cite{olcsva98}, introduce
dimensionless parameters
\begin{eqnarray}
 \tilde J&=&\frac{1}{MR^{2m}}\int_0^R \rho r^{2m+2} \mathrm d r\approx 1.64\times 10^{-2}, \label{tild_J} \\
 \tilde I&=&\frac{I}{MR^2}=\frac{8\pi}{3MR^2}\int_0^R \rho r^4 \mathrm d r\approx 0.261,  \label{tild_I}  \\
 \tilde Q&=&\frac{m(m+1)\tilde J}{4\tilde I}\approx 9.38\times 10^{-2}.  \label{tild_Q}
%, moment of inertia of the star
\end{eqnarray}

Equations (\ref{dot_Om})-(\ref{dot_T}) can be applied to model evolution of r-mode unstable star within slow rotating Newtonian approximation. As discussed in Section \ref{Sec_EvolSimple}, on the long timescales, evolution can be described by one equation  (\ref{Evol}), and for slow rotating Newtonian stellar models it can be rewritten in the form
\begin{equation}
  L_\mathrm{cool}=Q_\mathrm{other}+\frac{I\Omega}{3}\left|\dot \Omega^\mathrm{GW}\right|
  =Q_\mathrm{other}+\frac{1}{3}\left|\dot E_\mathrm{rot}^\mathrm{GW}\right|.
  \label{longTerm_Newton}
\end{equation}
The latter equality clearly demonstrates the budget of r-mode instability in slow rotating Newtonian model: $1/3$ of the GW spin down power  is converted to the heating (and finally emitted from the star by the cooling processes), and the remaining $2/3$ of $\left|\dot E_\mathrm{rot}^\mathrm{GW}\right|$ is directly emitted in form of GWs.

\subsection{Comparison with previous works} \label{Sec_Compar}
As discussed in Paper I, the evolution equations (\ref{dot_Om}-\ref{dot_T}) formally differ from ``standard'' equations derived by \cite{olcsva98,hl00}, which are widely applied  in the papers on the evolution of r-mode unstable NSs. Namely, in the leading order in mode amplitude, the latter equations can be written in form:
\begin{eqnarray}
\dot {\hat \Omega} &=&-\frac{2\tilde Q \alpha^2}{ \tau^\mathrm{dis}}\hat\Omega \label{dot_Om_stand}. \\
\dot \alpha &=& -\left(\frac{1}{\tau^\mathrm{GW}(\hat \Omega)}+\frac{1}{\tau^\mathrm{dis}}\right) \alpha \label{dot_alpha_stand}, \\
C\dot T &=& \frac{\tilde J M R^2\hat \Omega}{\tau^\mathrm{dis}} \alpha^2+
Q_\mathrm{other}-L_\mathrm{cool}. \label{dot_T_stand}
\end{eqnarray}

The difference occurs in spin frequency evolution equations (compare Eqs.\ \ref{dot_Om} and \ref{dot_Om_stand}). It originates from uncertainties in definition of mean spin frequency, caused by differential rotation, which can be generated in a star as a result of CFS instability (see, e.g., \citealt{Spruit99_DifRot,Rezzolla_etal00,lu01b,fll16} and Section \ref{Sec_DifRot}).
Namely, \cite{olcsva98,hl00} write total angular momentum as
\begin{equation}
J=I\hat \Omega+J_\mathrm{c}, \label{J_olcsva}
\end{equation}
where canonical angular momentum of r-mode is introduced:
\begin{equation}
J_\mathrm{c}=-(3/2)\hat\Omega\tilde J MR^2 \alpha^2.
\end{equation}
As a result, their spin frequency parameter is
\begin{equation}
\hat \Omega= (1+\tilde Q\alpha^2)\Omega. \label{Om_hat}
\end{equation}
After this quantity is introduced into standard evolution equations (Eqs.\ \ref{dot_Om_stand}-\ref{dot_T_stand})
they coincide with  Eqs.\ (\ref{dot_Om_stand}-\ref{dot_T_stand}) in the leading order in $\alpha$.

Note, for non-dissipative  stellar model ($\tau^\mathrm{dis}=\infty$) $\dot {\hat{\Omega}}=0$, but $\dot \Omega$ is finite and negative. In Section \ref{Sec_DifRotFll} I demonstrate that  nondissipative star should be observed as spinning down at rate $\dot \Omega_\mathrm{app}\sim (0.5\div 0.8)\,\dot \Omega$,  which only numerically differs from $\dot \Omega$.

Equations derived by \cite{olcsva98,hl00} contain the terms, which are of the next order in $\alpha^2$. As discussed in Paper I, these terms are not yet derived accurately (and thus are different for \citealt{olcsva98} and \citealt{hl00}), so it would be more self-consistent to omit them.

Equations suggested by \cite{hl00} contain additional term, which describes evolution of r-mode amplitude governed by the magnetic braking. Namely,
$\alpha/2\tau^\mathrm{M}$ is added to the right-hand side of Eq.\ (\ref{dot_alpha_stand}). This term is supported by the statement that canonical angular momentum is proportional to wave action, being thus adiabatic invariant. As discussed in Section \ref{Sec_DifRotGen}, similar term can be reproduced by the theory developed in Section \ref{Sec_GenEquations}, but it depends on differential rotation $\Delta \Omega$. In a particular case, when  differential motion of fluid elements is suppressed, e.g.\ by magnetic field (see Section \ref{Sec_DifMag}),  $\Delta \Omega$ is  unique (\citealt{Chugunov15}) and given by Eq.\ (\ref{Del_Om_nondrift}). In this case, assuming that all $\dot E_\mathrm{ex}^\mathrm{M}$ comes to the r-mode, Eq.\ (\ref{dotE_mag}) leads to additional term $\alpha/2\tau^\mathrm{M}$ in the right hand side of Eq.\ (\ref{dot_alpha}) and agrees with \cite{hl00}.

However, in a general case  $\Delta \Omega$ differs from Eq.\ (\ref{Del_Om_nondrift}) leading  to different additional term in Eq.\ (\ref{dot_alpha}), which contradicts \cite{hl00}. It can indicate that part of $\dot E_\mathrm{ex}^\mathrm{M}$ transfers not to the r-mode, but to the differential rotation. In this case,  differential rotation is enhanced by braking and becomes additional degree of freedom, controlled by external torque (as it is in the absence of oscillation modes). So, accurate treatment of CFS instability under action of external torque should deal with this feature. As far as I am concerned, such analysis has not been done yet.%

Hopefully, this uncertainty should not affect most applications of r-mode instability, because the magnetic braking terms
{in mode amplitude evolution} can be neglected if $\tau^\mathrm{M}\gg \tau^\mathrm{GW}$, which is typically the case.

\section{Role of differential rotation} \label{Sec_DifRot}
Equations  (\ref{dot_Om}-\ref{dot_T}) can be applied to study evolution of the mode amplitude $\alpha$, $\Omega$ , and $T$. However, as discussed in Section \ref{Sec_DifRotGen}, the differential rotation, associated with CFS instability, can affect the apparent frequency.
In this Section I analyse several simplified models, in which differential rotation can be treated accurately, and use these results to discuss the evolution of NS during recycling in LMXB.

\subsection{CFS instability and differential rotation in perfect fluid Newtonian stellar models}
\label{Sec_DifRotFll}

\cite{fll16} provides analytical solution for unstable r-modes  driven by gravitational radiation reaction force in perfect fluid Newtonian stellar model.
In particular, for the model with uniform slow rotation at spin frequency $\Omega_0$  at the initial moment of time they demonstrate that:

(a) r-mode grows exponentially and the first order velocity perturbation for r-mode with multipolarity $m$ can be expressed as [see Eqs.\ (77) and (93) by \cite{fll16}  and substitute $l=m$]:
\begin{equation}
  \delta^{(1)} \bm v=
   \mathfrak{Im}
   \left[
 \frac{\alpha_\mathrm{FLL}\, \Omega r^m}{m\,R^{m-1}}\bm r
       \times \nabla
         \left(
            \sin^m\theta\, \mathrm{e}^{i(m\phi+\omega t)}
         \right)
   \right]
   \,\mathrm e^{-t/\tau^\mathrm{GW}}.
 \label{vprofile_fll}
\end{equation}
Here $\omega$ is given by Eq.\ (\ref{om_r}) for $\Omega=\Omega_0$.
As noted above, the latter
agrees with Eq.\ (\ref{tau_GW}) for $\Omega=\Omega_0$. Note,
the dimensionless amplitude $\alpha_\mathrm{FLL}$
introduced by \cite{fll16} stays constant, while r-mode is growing.
In contrast, here (as in the most of the papers on r-mode
instability) the first order perturbation is described by `instantaneous' amplitude $\alpha=\alpha(t)$,
which determines amplitude of the first order velocity perturbations at
a given moment $t$
 (see, e.g., Eq.\ \ref{delta_v}). These amplitudes are
coupled according to
\begin{equation}
    \alpha^2_\mathrm{FLL}e^{-2t/\tau^\mathrm{GW}}=\frac{ (2m+1)!}{ 2^{2m+1}\,\pi\, m(m+1)[(m-1)!]^2} \alpha^2(t)
  %  =\frac{5}{8\pi}\alpha^2.
   \label{renorm}
\end{equation}
For $m=2$ one obtains $\alpha^2_\mathrm{FLL}e^{-2t/\tau^\mathrm{GW}}=5\alpha^2(t)/(8\pi)$.
As far as the results by \cite{fll16} were obtained within perturbation theory and do not include any saturation mechanism,  they can be directly applied only during limited time, i.e. as long as $\alpha(t)\ll 1$.

(b) the enhancement of r-mode is accompanied by enhancement
of oscillation averaged Eulerian perturbation of velocity, which is
axisymmetric and can be written in the form [see Eqs.\ (115),
(118) and (119) by \cite{fll16}]:
\begin{eqnarray}
 \delta^{(2)}_\mathrm{sym}\bm v&=&-\alpha_\mathrm{FLL}^2\exp(-2 t/\tau^\mathrm{GW})
 \Omega \varpi
 \left(\frac{\varpi}{R}\right)^{2m-4}
\label{v_sym}
\\
 &\times&\left[\frac{(m+1)^2}{8}\frac{\varpi^2}{R^2}
       +\frac{m^2-1}{2}\left(\Upsilon(\varpi)-\frac{z^2}{R^2}\right)\right]
       \bm e_\phi.
  \nonumber
\end{eqnarray}
Here $\bm e_\phi$  is a unit vector in the direction of $\phi$ and auxiliary function
\begin{equation}
 \Upsilon(\varpi)=
 \frac{
       \int_{-z_\mathrm s}^{z_\mathrm s} \rho z^2\mathrm d z
      }
      {R^2
       \int_{-z_\mathrm s}^{z_\mathrm s} \rho\, \mathrm d z
      }.
\end{equation}

The secular motion of
fluid elements (with respect to unperturbed uniform rotation at $\Omega_0$), can be described by the drift velocity $\delta^{(2)}_\mathrm{sec}\bm v$, which differs from $\delta^{(2)}_\mathrm{sym} v$
because of the Stokes drift (see, e.g., \citealt{Stokes1847,LH53}). For $m=2$ r-mode it  can be described by effective velocity (e.g., \citealt{Rezzolla_etal01a,Chugunov15,fll16})
\begin{eqnarray}
\bm v_\mathrm{S}=
 \frac{3\Omega \varpi}{4}\alpha_\mathrm{FLL}^2 \exp(-2t/\tau^\mathrm{GW})\left[\left(\frac{\varpi}{R}\right)^2-2\left(\frac{z}{R}\right)^2\right] \bm
e_\phi. \label{Stokes}
\end{eqnarray}
 As as result (\citealt{fll16}),
\begin{eqnarray}
 \delta^{(2)}_\mathrm{sec}\bm v&=& \delta^{(2)}_\mathrm{sym}\bm
 v+\bm v_\mathrm{S}
  \label{v_drift} %\nonumber % \label{dOmega}
 \\
 &=&
 -\frac{3}{2}\alpha_\mathrm{FLL}^2\exp(-2t/\tau^\mathrm{GW})
 \Omega \varpi
 \left[\frac{1}{4}\frac{\varpi^2}{R^2}
 +\Upsilon(\varpi)\right]
 \bm e_\phi.
 \nonumber
 %\label{v_drift}
\end{eqnarray}
where the last equality corresponds to $m=2$.
The resulting
motion of fluid elements (\ref{v_drift}) is axisymmetric,
cylindrically stratified (i.e., drift velocity depends
on $\varpi$ only), but differs from uniform rotation.

Let me now map these results to the variables introduced in this paper.
To obtain $\Omega$ as defined by Eq.\ (\ref{Omega}), I need to write down angular momentum as a function of time. As discussed  by \cite{lu01b} (see also Section \ref{Sec_slow}), only axisymmetric part of the velocity perturbations contribute to the total
angular momentum, which takes the form
\begin{eqnarray}
 \bm J(t)
&=&
\int \rho \left [\bm v_0\times \bm r\right] \mathrm d^3
+\int \rho \left [\delta^{(2)}_\mathrm{sym}  \bm v\times \bm r\right] \mathrm d^3
r
\nonumber \\
&=&
%=
\bm J_0-\frac{m(m+1)}{4}\tilde J MR^2\alpha^2 \Omega_0  \bm e_z,
\label{J_fll}
\end{eqnarray}
where Eqs.\ (\ref{tild_J}) and (\ref{renorm}) were applied.%
\footnote{This result can be also confirmed by calculating total torque associated with gravitational radiation-reaction force (see \cite{fll16} for explicit formulae for the latter force).}
It is worth to note, that the change of angular momentum $\bm J-\bm J_0$ is equal to the canonical angular momentum of r-mode $J_\mathrm{c}$ as defined by \cite{fs78a, fs78b},
as it should be for the eigenmodes evolving under the action of reaction-reaction force in absence of viscosity (\citealt{fs78a,fs78b}).%
\footnote{It is easy to show by explicit calculation that the same is true for the energy: change of the energy, which can be calculated by Eq.\ (\ref{E_slow_gen}), is equal to the canonical energy of r-mode $E_\mathrm{c}$, as it should be (\citealt{fs78a,fs78b}).}
Eq.\ (\ref{J_fll}) allows writing down the spin frequency $\Omega$ as defined in this paper
\begin{equation}
 \Omega(t)=\frac{J(t)}{I}=[1-\tilde Q \alpha^2(t)] \Omega_0
\label{Omega_t}
\end{equation}
Combined with Eq.\ (\ref{renorm}), it confirms Eqs.\ (\ref{dot_Om}) and (\ref{dot_alpha}) in the leading order in $\alpha$ (see Paper I for discussion of next-to-leading order corrections). %
The thermal evolution equation (Eq.\ \ref{dot_T}) is not affected by instability because dissipation is neglected in this subsection.

%%%%%%%%%%%%%%%%%%%%%%%%%%%%%%%%%%%%%%%%%%%%%%%%%%%%%%%%%%%%%%%%%%%%%
\begin{figure}
    \includegraphics[width=\columnwidth]{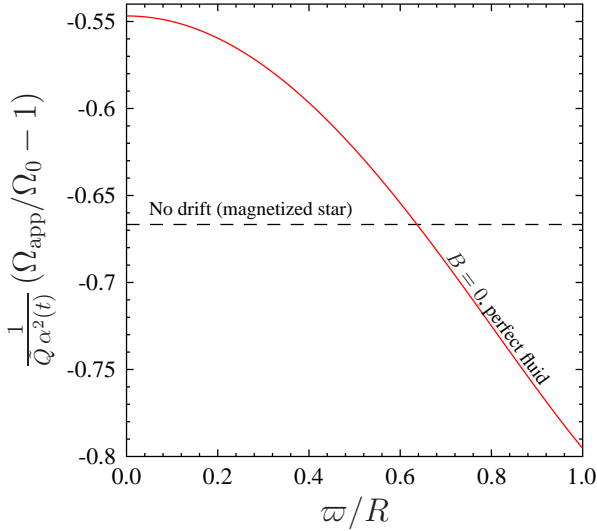}
    \caption{
        {
   Apparent change of spin frequency $\Omega_\mathrm{app}-\Omega_0$
    as function of cylindrical radius $\varpi$ of the observed point on the NS surface. Solid line corresponds to profile generated by r-mode instability in nonmagnetized perfect fluid, while dashed line is for magnetized Newtonian star}. The profile is normalized to $\Omega_0-\Omega(t)=\tilde Q\alpha^2(t) \Omega_0$.}
    \label{Fig_DifRotFll}
\end{figure}
%%%%%%%%%%%%%%%%%%%%%%%%%%%%%%%%%%%%%%%%%%%%%%%%%%%%%%%%%%%%%%%%%%%%%%

As noted above, the apparent spin frequency is typically associated with secular motion of  the specific point at the stellar surface, and thus it is given by
\begin{equation}
    \Omega_\mathrm{app}=\Omega_0+\frac{\delta^{(2)}_\mathrm{sec}\bm v}{\varpi}
    =\left\{1-
        \frac{15\,\alpha^2(t)}{16\pi}
    \left[\frac{\varpi^2}{4R^2}
    +\Upsilon(\varpi)\right]
    \right\}\Omega_0. \label{Om_obs_fll}
\end{equation}
It leads to apparent spin-down rate
\begin{equation}
\frac{\mathrm d \Omega_\mathrm{app}}{\mathrm d t}=\frac{15\,\alpha^2(t)}{8\pi\tau^\mathrm{GW}}
\left[\frac{\varpi^2}{4R^2}
+\Upsilon(\varpi)\right]
\Omega_0, \label{dOm_obs_fll}
\end{equation}
which depends on $\varpi$ -- {cylindrical radius} of the observed point at the surface.

Corresponding profile of $\Omega_\mathrm{app}$
for perfect fluid Newtonian star with polytropic equation of state with $n=1$  is shown by solid line in  Fig.\ \ref{Fig_DifRotFll}.   $\Omega_\mathrm{app}$   differs from $\Omega(t)$,
given by the Eq.\ (\ref{Omega_t}),
even in the leading order in $\alpha(t)$.  This results agrees with general arguments presented at Section \ref{Sec_DifRotGen} because $\dot E_\mathrm{ex}=\dot E_\mathrm{ex}^\mathrm{GW}$ in the absence of dissipation (see Eq.\ \ref{Eex_evol_gen}).
The
 profile
\begin{equation}
\Delta  \Omega=\Omega_\mathrm{app}-\Omega
=\left\{\tilde{Q}-
\frac{15}{16\pi}
\left[\frac{\varpi^2}{4R^2}
+\Upsilon(\varpi)\right]
\right\}\alpha^2(t)\Omega_0. \label{DeltaOm_fll}
\end{equation}
evolves on the gravitational instability timescale.

Summarizing, for non-magnetized  Newtonian models of a perfect-fluid star, development of the r-mode instability agrees with evolution equations (\ref{dot_Om})--(\ref{dot_T}),
however the apparent spin down rate is
affected by differential rotation (see Eq. \ref{dOm_obs_fll}).
It is worth to note, that  the GW emission associated with CFS instability results in spin down of all points at stellar surface.%
\footnote{Note, Fig.\ 1 in \cite{fll16} represents axisymmetric  Eulerian perturbation of the velocity, which indeed can be positive at certain part of the stellar surface,
but differs from secular motion of fluid elements due to Stokes drift (see e.g.\ \citealt{Stokes1847,LH53}).}
In particular, it demonstrates that standard equation  (\ref{dot_Om_stand}),  predicting $\hat \Omega=\mathrm{const}$ for absence of dissipation ($\tau^\mathrm{dis}=0$), can be rather misleading, if we do not take into account differential rotation effects (i.e. treat $\hat \Omega$ as apparent frequency).

\subsubsection{Saturated r-modes in non-magnetized perfect fluid stellar models}\label{Sec_SaturNonmagId}
Previous discussion
holds true for nonsaturated modes.
For nondissipative star mode coupling mechanism of saturation (\citealt{btw04a, btw04b, btw09, bw13}) can be suppressed because vanishing damping rate of daughter modes require almost perfect tuning in mode triplet (see, e.g., Eq.\ 4 in \citealt{bw13}). Indeed, as shown by \cite{Kastaun11}, in the absence of dissipation r-modes can exist with very large amplitudes $\alpha\sim 1$.
However, below I apply traditional simplified model of saturation (e.g.\ \citealt{levin99}) and assume mode amplitude to be constant after it reaches saturation value $\alpha^\mathrm{sat}\ll1$.
In this case the evolution of differential rotation can be qualitatively described by the toy model by \cite{flrc17} at magnetic field $B\rightarrow0$ (see also Section \ref{Sec_DifVisc}). The model discusses incompressible perfectly conducting rotating fluid with cylindrical symmetry, driven by a force, which mimics the radiation-reaction force associated with CFS instability. The amplitude of this force (per unit volume) is $f^\mathrm{GW}\sim f(\varpi)\alpha^2\rho R\Omega/\tau^\mathrm{GW}$. {Here $f(\varpi)$ encodes dependence of $f^\mathrm{GW}$ on $\varpi$.} In absence of viscosity and magnetic field, rotation of cylindrical layers is uncoupled. As a result,  constant force $f^\mathrm{GW}$ leads to linear increase of differential rotation with time at rate $\Delta \dot \Omega\sim  f (\varpi) \alpha^2 \Omega /\tau^\mathrm{GW}$,
which depends on the  {cylindrical radius and contributes to apparent spin down}. Thus the differential rotation affects apparent spin-down for saturated r-modes in nonmagnetized perfect fluid stellar models.

\subsection{Differential rotation in viscous stellar models}
\label{Sec_DifVisc}

Previous discussion in this section neglects viscosity, which can affect profile of differential rotation even it is not enough strong to suppress instability as it is.
In this subsection I account for these effects qualitatively within a bit modified toy model by \cite{flrc17}: I neglect the magnetic field, but account for viscous terms.
As described in Subsection \ref{Sec_SaturNonmagId}, the model discusses incompressible rotating fluid with cylindrical symmetry, driven by a force $f^\mathrm{GW}$, which mimics the radiation-reaction force associated with CFS instability. The only nonvanishing component of the velocity is $v_\phi$; it is governed by equation (e.g., \citealt{ll87})
\begin{equation}
  \rho\frac{\partial v_\phi}{\partial t}
  +\eta\left(
     \frac{1}{r}\frac{\partial}{\partial r}r  \frac{\partial v_\phi}{\partial r}-\frac{v_\phi}{r^2}\right)=f^\mathrm{GW}. \label{DifRotVisc_Gen}
\end{equation}
In the absence of external force, general solution of this equation can be presented as linear combination of uniform rotation and infinite set of exponentially decaying ($\propto \exp(-\gamma_i t)$, where $\gamma_i>0$) eigenfunctions, with radial dependence $\propto I_1(\sqrt{\rho\gamma_i/\eta}r)$. Here $I_1$ is modified Bessel function of the first kind (solution with modified Bessel function of the second kind are not regular at $r\rightarrow 0$ and thus unphysical).
Exact value of time-scales $\gamma_i$ depends on the condition at the external boundary of the cylinder, but subsequently I will need only order-of-magnitude estimate: $\gamma_i\sim \gamma\equiv\eta/(\rho\, R^2)$.
Note, that this estimate agrees by order-of-magnitude the with shear viscosity dissipation timescale $\tau^\mathrm{S}$, given by Eq.\ (\ref{tau_S}).
Below I examine the differential rotation before and after saturation of r-modes.

\subsubsection{Unsaturated r-modes}
According to Eq.\ (\ref{dot_alpha}), the mode amplitude is exponentially increasing at timescale $\tau$ given by:
\begin{equation}
\frac{1}{\tau}=\frac{1}{\tau^\mathrm{GW}}+\frac{1}{\tau^\mathrm{dis}}.
\end{equation}
Here I discuss evolution of differential rotation for unstable star {($\tau<0$)} and neglect evolution of $\Omega$ and $T$ in equations (\ref{dot_Om}-\ref{dot_T}) for simplicity Thus, I treat $\tau$ as a constant.

The radiation reaction force can be estimated as (\citealt{fll16})
\begin{equation}
 f^\mathrm{GW}\sim -\alpha^2(t)\rho R\frac{\Omega}{\tau^\mathrm{GW}}=
  -\alpha^2(0)\exp\left(-\frac{2t}{\tau}\right)\rho R\frac{\Omega}{\tau^\mathrm{GW}}.
 \label{fGR_visc_unsat}
\end{equation}
{The minus sign demonstrates that $f^\mathrm{GW}$ brakes NS rotation.}
Having expanded this force over eigenfunctions of Eq.\ (\ref{DifRotVisc_Gen}), I got the following estimate for %differential rotation
\begin{equation}
\Delta \Omega\sim -\frac{\alpha^2(t)}{\gamma \tau-2}\frac{\tau}{\tau_\mathrm{GW}} \Omega.
\end{equation}
The contribution of $\Delta \Omega$ to the apparent spin down can be estimated as
\begin{equation}
 \frac{\mathrm d \Delta \Omega}{\mathrm d t}\sim -\frac{2}{\tau} \Delta \Omega\sim
  \frac{\alpha^2(t)}{\gamma \tau-2}\frac{2}{\tau_\mathrm{GW}} \Omega.
\end{equation}
This value is of the same order of magnitude as $\dot \Omega$, and thus differential rotation can affect apparent spin down rate for viscous stellar models.

\subsubsection{Saturated r-modes}\label{Sec_DifViscSatur}

To make the discussion more simple, I introduce several evolution timescales
\begin{eqnarray}
\beta^{\Delta \Omega}\equiv\frac{1}{\tau^{\Delta \Omega}}&\equiv&   \frac{1}{\Delta \Omega}
\frac{\mathrm d \Delta \Omega}{\mathrm d t}\\
\beta^{\Omega}\equiv\frac{1}{\tau^{\Omega}}&\equiv& \frac{1}{\Omega}
\frac{\mathrm d \Omega}{\mathrm d t}\\
\beta^{\alpha}\equiv\frac{1}{\tau^{\alpha}}&\equiv& \frac{1}{\alpha}
\frac{\mathrm d \alpha}{\mathrm d t}
\end{eqnarray}

A simplified treatment of saturation (\citealt{olcsva98,hl00,as14,gck14a}) is substitution of $\tau^\mathrm{dis}$ with the effective nonlinear dissipation timescale  $\tau^\mathrm{dis}_\mathrm{eff}=\left|\tau^\mathrm{GW}\right|$ in the whole equations (\ref{dot_Om_stand})--(\ref{dot_T_stand}). As a result, we have $\alpha(t)=\alpha^\mathrm{sat}$ and, within the toy model, time-independent
\begin{equation}
f^\mathrm{GW}\sim -(\alpha^\mathrm{sat})^2\rho R\frac{\Omega}{\tau^\mathrm{GW}}.
\end{equation}
As discussed in Section \ref{Sec_SaturNonmagId},
for non-magnetized perfect fluid stellar models differential rotation is linearly increasing with time and thus $\Delta \Omega$ can strongly exceed  $\alpha^2\Omega$ for $t\gg \tau^\mathrm{GW}$.
However, shear viscosity $\eta$ limits the growth of differential rotation.
Limiting amplitude of $\Delta \Omega$ can be estimated by  equating typical viscous force $f^\eta\sim \eta \Delta\Omega/R$ to $f^\mathrm{GW}$.
It leads to
\begin{equation}
\Delta \Omega\sim
\left(R\,\alpha^\mathrm{sat}\right)^2  \frac{\rho}{\eta}\frac{\Omega}{\tau^\mathrm{GW}}
\sim  \left(\alpha^\mathrm{sat}\right)^2 \frac{\Omega}{\gamma\,\tau^\mathrm{GW}}.\label{DelOm_shear}
\end{equation}
This result can be confirmed by accurate expansion of $f^\mathrm{GW}$ over eigenfunctions of Eq.\ (\ref{DifRotVisc_Gen}).
As far as $\gamma^{-1}\sim \tau^\mathrm{S}$ and for unstable star $\tau^\mathrm{S}>\left|\tau^\mathrm{GW}\right|$,
$\Delta \Omega$ can exceed $\alpha^2\Omega$ for a factor $\sim 1/\left|\gamma\,\tau^\mathrm{GW}\right|\sim \tau^\mathrm{S}/\left|\tau^\mathrm{GW}\right|$.

As long as $\alpha^\mathrm{sat}$ depends on  $\Omega$ and $T$, $\Delta \Omega$ can vary on the same timescale as $\Omega$ and $T$, $\tau^\mathrm{\Omega}\sim\tau^\mathrm{GW}/\alpha^{2}$ (see Eq.\ \ref{dot_Om}),%
\footnote{As discussed in Section \ref{Sec_longEvol}, a star with saturated unstable mode evolves along thermal equilibrium curve on $(\Omega,T)$ plane, thus the timescale of temperature changes is the same as $\tau^\mathrm{\Omega}$.}
which is very long in comparison with $\tau^\mathrm{GW}$
and cannot contribute to the apparent spin down as long as $\Delta \Omega\ll \Omega$.

More accurate treatment of the saturation within the mode coupling mechanism (e.g.\ \citealt{btw09}) demonstrates that in some cases the r-mode amplitude is modulated and oscillates near saturation value due to complex behaviour of the mode coupling.
It can lead to modulation of the $\Delta \Omega$. To model this effect I consider Eq.\ (\ref{DifRotVisc_Gen}) with oscillating  external force $f^\mathrm{GW}\sim \alpha^2\rho R\Omega\,[1+\sin(t \beta^\alpha)]/\tau^\mathrm{GW}$ in the right part.
Expanding this force over eigenmodes of Eq.\ (\ref{DifRotVisc_Gen}), I estimate the typical $\Delta \Omega$ rate of change as
\begin{equation}
 \frac{\mathrm d \Delta \Omega}{\mathrm d t}\sim
\frac{\beta^\alpha }{\sqrt{\left(\beta^\alpha\right)^2+\gamma^2}}
\frac{\mathrm d \Omega}{\mathrm d t}, \label{SpinDownSaturVisc}
\end{equation}
where Eq.\ (\ref{dot_Om}) was applied.
It means, that for $\beta^\alpha\gtrsim \gamma$ differential rotation can contribute to the apparent spin down. This case is not excluded in realistic models.
For example, let us consider a model shown in Fig.\ 4b by  \cite{btw09}.
The modulation timescale (estimated by sight from the plot) is  $\tau^\alpha\lesssim 10^6$~s. It is short in comparison with  $\gamma\sim\tau^\mathrm{S}$ for the temperature of the star  discussed on that plot (see Eq.\ \ref{tau_S} for $T\sim 10^{9-10}$~K). Thus, the apparent spin down can be affected. However,
the effect of modulation should be smoothed out on timescales much longer than $\tau^\alpha$. Thus, even in the considered example, the spin down rate, which is estimated from the variation of apparent frequency on the observational timescale $\tau^\mathrm{obs}\sim 1\mathrm{~yr}\gg \tau^\alpha$, should not be affected strongly.

The latter argument can be generalized: the modulation of the {apparent} spin down of a star by differential rotation can be {deduced from observations} only if one can measure it on the timescale $\tau^\mathrm{obs}\sim \tau^\alpha$.  Note, $\tau^\alpha\sim 1/\gamma$ gives an upper limit for a timescale allowed for measuring the effect of differential rotation on spin down, because for significantly longer  $\tau^\alpha\gg 1/\gamma$, contribution of differential rotation to the apparent spin down   is suppressed  for a factor $1/\gamma\tau^\alpha$ (Eq.\ \ref{SpinDownSaturVisc}).

Is it realistic to measure spin down at such timescale?
Indeed, variation of apparent frequency can hardly be measured with high precision, if spin down does not result in the shift of rotation phase $\gtrsim \pi$.
This requirement provides a lower bound for $\left|\dot \Omega\right|\gtrsim \pi/\left(\tau^\mathrm{obs}\right)^2$, corresponding to lower bound for $\alpha^\mathrm{sat}$, which may allow spin down measurement at time-scale $\tau^\mathrm{obs}$
\begin{eqnarray}
\alpha^\mathrm{sat}&\gtrsim& \sqrt{\frac{\tau^\mathrm{GW}}{2\nu\tilde Q (\tau^\mathrm{obs})^2}}
\approx  1.4\times10^{-5} \left(\frac{T}{10^8~\mathrm{K}}\right)^{2} \label{al_dif_rot_mes} \\
&\times&
 \left(\frac{R}{10\mathrm{\ km}}\right)^{3}
\left(\frac{M}{1.4M_\odot}\right)^{-3/2}
\left(\frac{\nu}{600\mathrm{\ Hz}}\right)^{-7/2}.
\nonumber
\end{eqnarray}
The latter estimate corresponds to $\tau^\mathrm{obs}=\tau^\mathrm{S}$.

Summarizing, in principle differential rotation can affect apparent spin down in viscous non-magnetized star. To do so, the mode amplitude should vary on timescale $\tau^\alpha\lesssim 1/\gamma$ and satisfy condition (\ref{al_dif_rot_mes}) for $\tau^\mathrm{obs}\sim \tau^\alpha$.

\subsection{Differential rotation in magnetized star} \label{Sec_DifMag}

Now I examine the role of differential rotation in magnetized star.
Differential motion of the fluid elements is the reason for magnetic field bending, which produces back reaction: magnetic stresses, generated by the bending. These stresses, in turn, act against subsequent bending and, if the velocity of the fluid motion is below the Alf\'en velocity, the magnetic force line have enough time to straighten (e.g.\ Section 3.14.2 by \citealt{AF63}), thus preventing the secular motion of the fluid elements (see, e.g., appendix B of \cite{Rezzolla_etal01a} for explicit equation, which couples fluid elements motion and magnetic field evolution).
Indeed, as shown by \cite{Chugunov15,flrc17}, in a particular case of r-modes in {prefect fluid} Newtonian stellar models, the relative displacement of the fluid elements is strongly suppressed, if the magnetic field $B$ is not too weak (namely, if the typical Alfv\'en timescale $\tau_\mathrm{A}\sim R\sqrt{\rho}/B$ does not exceed gravitational instability timescale, corresponding to {internal magnetic field} $B\gtrsim 3\times 10^8$~G for non-superconducting NS core;%
\footnote{{In is worth to stress that the toy model by \cite{flrc17} allows straightforward generalization for r-mode instability in dissipative star, by substitution $\tau^\alpha$ instead of $\tau_\mathrm{GW}$ as an amplitude  growing rate. It leads to conclusion that relative motion of fluid elements is suppressed if $\tau_\mathrm{A}\lesssim \tau^\alpha$. The latter condition is equal to $B\gtrsim 10^6$~G for $\tau^\alpha\sim 1$~yr. If it is satisfied, results of this section holds true.
\label{footnote_mag} }
}
subsequently I  assume that this condition is true for magnetized star).%
\footnote{As argued by \cite{Chugunov15} and proved by \cite{flrc17} the enhancement of the magnetic field by differential motion of fluid elements does not lead to instability suppression, at least for low saturation amplitudes. The first order perturbation, given by Eq.\ (\ref{vprofile_fll}), are almost unaffected by magnetic field up to $B\lesssim 10^{16}$~G (e.g., \citealt{Lee05,arr12,cs13,aly15}).}
As a result secular motion of fluid elements becomes close to uniform rotation. The axisymmetric second order Eulerian perturbation of the velocity $\delta^{(2)} \bm v_\mathrm{sym}$ becomes:
\begin{eqnarray}
\delta^{(2)} \bm v_\mathrm{sym}&=&-\bm v_\mathrm{S} +[\delta \bm \Omega\times \bm r] \label{deltaV_mag}
\\
&=&\frac{15\Omega \varpi}{32}\alpha^2(t)\left[2\left(\frac{z}{R}\right)^2-\left(\frac{\varpi}{R}\right)^2\right] \bm
e_\phi+\delta \Omega \varpi\bm
e_\phi,
\nonumber
\end{eqnarray}
which differs from uniform rotation
due to contribution of the Stokes drift to the secular motion of the fluid elements (see Eq.\ \ref{v_drift} and  \citealt{Chugunov15} for more details).
The $\delta \bm \Omega$ term is required to keep $\delta J=0$ according to definition of perturbations in this paper, which leads to  (see Eq.\ \ref{J_constraint}):
\begin{equation}
\delta \Omega=\frac{\tilde J}{2\tilde I}\alpha^2\Omega=\frac{\tilde Q}{3}\alpha^2\Omega. \label{Del_Om_nondrift}
\end{equation}
Note, in spite of the fact that the secular motion of the fluid elements is uniform rotation, spin frequency of this rotation (and the apparent spin frequency) differs from $\Omega$ because the drift velocity, given by  Eq.\ (\ref{Del_Om_nondrift}), is not vanishing. This effect can be described by
$\Delta \Omega=\delta \Omega$, which does not depend on the position {(cylindrical radius)} of the observed point at the stellar surface (see dashed line in Fig.\ \ref{Fig_DifRotFll}).
Correspondent apparent spin down rate is
\begin{equation}
\frac{\mathrm d \Omega_\mathrm{app}}{\mathrm d t}
=\frac{4\tilde Q \alpha^2}{3\tau^\mathrm{GW}(\Omega)}\Omega.\label{dOm_obs_mag}
\end{equation}
It is negative ($\tau^\mathrm{GW}<0$), and differs from
%$\mathrm d \Omega/ \mathrm d t$,
$\dot \Omega$, given by Eq. (\ref{dot_Om}), for a factor of $2/3$.
Note,  (\ref{dOm_obs_mag}) provides reasonable estimate for  $\dot \Omega_\mathrm{app}$ in perfect fluid non-magnetized star (compare solid and dashed lines in Fig.\ \ref{Fig_DifRotFll}).

\subsubsection{Saturated r-modes in magnetized star models}\label{Sec_DifMagSatur}
 As discussed by \cite{Chugunov15,flrc17} saturated modes requires special consideration.
Namely, as shown by \cite{flrc17} the magnetic field stops to grow soon after saturation and  secular motion of fluid elements becomes uniform rotation. As it was discussed in previous Section the requirement of $\delta J=0$ may lead to%
\footnote{$\Delta \Omega$ can differ from the value, given by Eq.\ (\ref{Del_Om_nondrift}) due to Stokes drift  with daughter modes, if the saturation occurs, produced by the mode coupling mechanism, as suggested by \citealt{arras_et_al_03,btw04a,btw04b,btw07,btw09}.}
\begin{equation}
\Delta \Omega\sim \alpha^2\Omega
\end{equation}
However, it is reasonable to assume that $\Delta \Omega$ depend on time at the same timescale as the mode amplitude $\alpha$, which is long compared to $\tau^\mathrm{GW}$ for saturated r-mode.
Thus, the contribution of $\Delta \Omega$ to the apparent spin down rate is $\sim \alpha^2 \dot \Omega$, which can be neglected for $\alpha\ll1$. Thus, one can directly confront $\dot \Omega$, given by Eq.\ (\ref{dot_Om}) with apparent spin-down rate.

\subsection{Differential rotation along recycling scenario}\label{Sec_DifRotRecyc}
According to recycling scenario of MSP formation, old slowly rotating NS is recycled by accretion from companion star and forms MSP or HOFNAR after accretion ends (\citealt{bkk76, acrs82,cgk14}).
Along this evolution,  NS can become unstable (e.g.\ \citealt{levin99}), but the star either evolves along the boundary of the instability window (e.g., \citealt*{ajk02}; \citealt{gck14a,gck14b,cgk17}) or plunges deeply into instability window with already saturated r-mode (see, e.g., discussion and analytic formulae in appendix B \citealt{gck14b}).
Note, as discussed by \cite{kgc16}, for a neutron star evolving along the boundary of the instability region the r-mode amplitude either is almost constant (and equal to the thermal equilibrium value $\alpha^\mathrm{eq}$) or oscillates, preserving root-mean-square value of amplitude equal to $\alpha^\mathrm{eq}$ (so called $\alpha$-oscillations). In both cases role of differential rotation in apparent spin rate can be described in the same way as it was done here for saturated modes (Secs.\ \ref{Sec_DifViscSatur}, \ref{Sec_DifMagSatur}).

Applying the results from these subsections, I come to a conclusion that differential rotation can affect apparent spin down frequencies only for non-magnetized star and only at short timescales $\lesssim \tau_\mathrm{S}$. {Note, that according to footnote \ref{footnote_mag}, the internal magnetic field should be $B\lesssim 10^6\mathrm{\,G}\,(\mathrm{1\,yr}/\tau^\alpha)$ to treat the star as non-magnetized. Typical surface magnetic field of  NSs in LMXBs and MSPs is much larger ($\sim 10^8-10^9$~G), thus if internal magnetic field is not strongly suppressed with respect to surface value, differential rotation does not affect their apparent spin down rate. Even if the field is suppressed,}
%Furthermore,
still, relatively high  r-mode amplitude (Eq.\ \ref{al_dif_rot_mes}) is required so that one could measure of the spin down rate.
I estimate the corresponding intrinsic GW strain
amplitude $h_0$  by using Eq.\ (23) by \cite{owen10}
\begin{eqnarray}
h_0&\gtrsim& 1.5\times10^{-25} \left(\frac{R}{10\,\mathrm{km}}\right)^{6}
\left(\frac{T}{10^8\,\mathrm{K}}\right)^{2}
\left(\frac{\nu}{600\,\mathrm{Hz}}\right)^{-1/2}
\nonumber\\
&\times&
\left(\frac{M}{1.4 M_\odot}\right)^{-1/2}
\left(\frac{r}{1\,\mathrm{kpc}}\right)^{-1},
\end{eqnarray}
Here $r$ is distance to the NS. This amplitude is a bit below sensitivity of continuous GW searches (e.g.\ \citealt{LIGO18_AllSky}) and thus hypothesis that r-modes with such amplitudes are excited in some of NSs cannot be rejected on the base of direct GW observations. However, indirect observational constraints are possible. Namely, for known neutron stars in LMXBs, amplitudes given by Eq.\ (\ref{al_dif_rot_mes}) require enhanced cooling (e.g., direct URCA neutrino emission in the core), so that these NS could to be in thermal equilibrium  (e.g.\ \citealt{ms13,kgc16}), and may be too large to be consistent with apparent spin down rates for some of the sources (see \citealt{ms13} for details). Hence, it is unlikely that differential rotation affects apparent spin down rate for NSs in LMXBs. The r-mode amplitude in {known} MSPs is constrained even stronger ($\alpha\lesssim 10^{-7}$, e.g.\ \citealt{as15_oscMSP,Schwenzer_etal_Xray,cgk17,47_Tuc_aa}) and thus differential rotation cannot affect the apparent spin down rate of {known} MSPs.

\section{Summary, conclusions, and caveats}\label{Sec_DisConc}

In this paper I discuss possible imprints  of differential rotation on the evolution of the apparent spin frequency $\Omega_\mathrm{app}$ of CFS unstable star.
Namely, if an unstable star emits GWs, differential rotation and angular momentum evolve. Even though differential rotation is weak, it can evolve much faster than the angular momentum and contribute to the apparent spin-down rate  $\dot \Omega_\mathrm{app}$ (see Section \ref{Sec_DifRotGen}). Additional problem, associated with  differential rotation, is the uncertainty of the effective spin frequency definition, the results of which is, in particular, different form of the effective spin down rates suggested by \cite{olcsva98,hl00} and paper I. Two former papers predict that spin down is associated with dissipation, but the latter couples spin down directly with instability timescale. As shown in paper I, this difference is associated only with definition of the effective spin frequency, but still there is a question: which value should corresponds to the apparent spin down rate?

In Section \ref{Sec_longEvol} I analyse the evolution on the long timescales ({total} change of the rotational energy due to GW emission strongly exceeds the energies of the unstable mode and the thermal energy).
I demonstrate that the evolution equations can be reduced to one differential equation, which describes evolution rate along the sequence of thermally equilibrium states. This equation can be applied not only to study evolution within certain model of r-mode instability suppression, but also to constrain the shape of the
instability windows and r-mode amplitude from
observational data. For Newtonian stellar models this equation have especially simple form (Eq.\ \ref{longTerm_Newton}), which was previously known (e.g.,\ \citealt{cgk14}). Differential rotation does not affect evolution on the long timescales.

Evolution on the short timescales require more detailed analysis.
The apparent frequency is typically associated with motion of certain part of stellar surface (hot spot, magnetic pole, etc.), and thus {even for given angular momentum} it generally depend on location of the observed part of the star and can be affected by differential rotation. In  Sections \ref{Sec_DifRotFll} and \ref{Sec_DifMag} I demonstrate that  these features indeed take place for non-dissipative stellar models (except for the saturated r-modes in magnetized star, where $\dot \Omega_\mathrm{app}=\dot \Omega$, given by Eq.\ \ref{dot_Om}). In particular, I demonstrate that the star should be observed as spinning down, contrary to the naive expectation, which directly applies equations by \cite{olcsva98,hl00} and predicts constant spin frequency for absence of dissipation.
However, realistic model of NS should be dissipative.
I discuss such models qualitatively in Section \ref{Sec_DifVisc} and formulate criteria for the case, when differential rotation can affect observational evolution of the spin frequency: the r-mode amplitude should vary faster than the shear viscosity dissipation timescale and it should be large enough {to produce spin down, which can be measured during this timescale}.

In Section \ref{Sec_DifRotRecyc}, I argue that it is unlikely that differential rotation can have an effect upon observations of NSs evolving along recycling scenario. Namely, detailed evolution scenarios (e.g., \citealt{gck14a,gck14b,cgk17}) predict that even if NS becomes unstable during its evolution, the instability is either saturated or almost suppressed by dissipation.

As discussed in Section \ref{Sec_DifViscSatur}, differential rotation, in principle, can affect apparent spin down rate for non-magnetized star.  However,
{even very small internal magnetic field $B\gtrsim 10^6 (\mathrm{1\,yr}/\tau^\alpha)$~G suppresses these effects and } indirect observational constraints to the r-mode amplitude  {keep differential rotation effects beyond measuring capabilities}.
Still, strictly speaking, {these effects} are not excluded in a newly born NS or even in a recycling neutron star, provided that observation of this particular NS allow of the large r-mode amplitude
{and negligible internal magnetic field}.

The main practical conclusion of the paper is the following:
$\Omega$ defined by Eq.\ (\ref{Omega}) is closer to the
secular motion of the fluid elements than $\hat \Omega$, suggested by \cite{olcsva98} (see Fig.\ \ref{Fig_DifRotFll}). It favours equations (\ref{dot_Om})-(\ref{dot_T}) from Paper I for the modelling of r-mode unstable NS.
However, as discussed in Section \ref{Sec_DifRotRecyc}, differential rotation does not affect NS recycling in realistic (dissipative) model. In particular, it means, that the rate of change of effective frequency, $\dot \Omega$ and $\dot{\hat{\Omega}}$, should be almost the same (at the leading order in $\alpha$)%
\footnote{This result can be obtained directly from Eqs.\ (\ref{dot_Om}) and (\ref{dot_Om_stand}) and the fact that either the unstable NS is located near boundary of instability region or r-mode is saturated (see appendix B of \citealt{gck14b}). In the first case, $\tau^\mathrm{dis}\approx -\tau^\mathrm{GW}$ by definition of the boundary of instability region, in the second case $\tau^\mathrm{dis}$ in Eq.\ (\ref{dot_Om_stand}) should be substituted to $\tau^\mathrm{dis}_\mathrm{eff}=-\tau^\mathrm{GW}$. In both cases Eq.\ (\ref{dot_Om_stand}) becomes equal to (\ref{dot_Om}).
}
and can be directly compared with the apparent spin down rate, as it was performed by \cite{kgc16,aaehh18}.
Note, however, that this conclusion does not holds true, if we consider strongly unstable star NS with initially small (non-saturated) r-mode amplitude (see Section \ref{Sec_DifRot}).

The above statements comes with some caveats, which, as I expect, do not affect general results. First, the discussion in Section \ref{Sec_DifRot} is based on Newtonian models, and, strictly speaking, conclusions' validity should be checked in a generally relativistic case. I also use very simplified microphysics (barotropic equation of state, with even polytropic approximation for numerical estimates). In particular, I neglect superfluidity in NS core, which can lead to high frequency g-modes (see, e.g., \citealt{kg14,dg16}). Although these assumptions are widely used in r-mode literature, their validity should be checked by more refined models.
I also neglect that contribution to the differential rotation, which are associated with external (non CFS) torques. Such differential rotation can affect observations (e.g., pulsar glitches are interpreted as effect of relaxation of differential motion of superfluid and normal components of the star, e.g., \citealt{ai75}). However, if the differential rotation induced by external torque is small, it can be (likely) decoupled from the effects discussed here. The strong differential rotation can affect frequencies of unstable oscillations (e.g., \citealt{csy14}), but these effects are left beyond scope of this paper. Finally, I do not discuss effects of the solid crust, which can have its own spin frequency, coinciding with the observed one.
{If crust is magnetically coupled with the core, differential motion of the crust with respect to fluid elements in the core should be suppressed and results of Sec.\ \ref{Sec_DifMag} should be generally applicable. In absence of such coupling, an obvious toy model of the crust motion can be obtained by assumption that the crustal spin frequency is given by some average over secular motion of the fluid elements on the top of the core. Within this model all results of the paper are also applicable. However, a strict proof of this statement goes beyond the scope of this paper.}

%%%%%%%%%%%%%%%%%%%%%%%%%%%%%%%%%%%%%%%%%%%%%%%%%%%%%%%
\section*{Acknowledgements}
%%%%%%%%%%%%%%%%%%%%%%%%%%%%%%%%%%%%%%%%%%%%%%%%%%%%%%%
{I thank anonymous referee for useful comments, which helped me to improve the paper}.

\label{lastpage}

\begin{thebibliography}{}
\makeatletter \relax
\def\mn@urlcharsother{\let\do\@makeother \do\$\do\&\do\#\do\^\do\_\do\%\do\~}
\def\mn@doi{\begingroup\mn@urlcharsother \@ifnextchar [ {\mn@doi@}
  {\mn@doi@[]}}
\def\mn@doi@[#1]#2{\def\@tempa{#1}\ifx\@tempa\@empty \href
  {http://dx.doi.org/#2} {doi:#2}\else \href {http://dx.doi.org/#2} {#1}\fi
  \endgroup}
\def\mn@eprint#1#2{\mn@eprint@#1:#2::\@nil}
\def\mn@eprint@arXiv#1{\href {http://arxiv.org/abs/#1} {{\tt arXiv:#1}}}
\def\mn@eprint@dblp#1{\href {http://dblp.uni-trier.de/rec/bibtex/#1.xml}
  {dblp:#1}}
\def\mn@eprint@#1:#2:#3:#4\@nil{\def\@tempa {#1}\def\@tempb {#2}\def\@tempc
  {#3}\ifx \@tempc \@empty \let \@tempc \@tempb \let \@tempb \@tempa \fi \ifx
  \@tempb \@empty \def\@tempb {arXiv}\fi \@ifundefined
  {mn@eprint@\@tempb}{\@tempb:\@tempc}{\expandafter \expandafter \csname
  mn@eprint@\@tempb\endcsname \expandafter{\@tempc}}}

\bibitem[\protect\citeauthoryear{{Abbassi}, {Rieutord}  \& {Rezania}}{{Abbassi}
  et~al.}{2012}]{arr12}
{Abbassi} S.,  {Rieutord} M.,   {Rezania} V.,  2012,
\mn@doi [\mnras]
  {10.1111/j.1365-2966.2011.19930.x}, \href
  {http://adsabs.harvard.edu/abs/2012MNRAS.419.2893A} {419, 2893}

\bibitem[\protect\citeauthoryear{{Akmal}, {Pandharipande}  \&
  {Ravenhall}}{{Akmal} et~al.}{1998}]{apr98}
{Akmal} A.,  {Pandharipande} V.~R.,   {Ravenhall} D.~G.,
1998, \mn@doi [\prc]
  {10.1103/PhysRevC.58.1804}, \href
  {http://adsabs.harvard.edu/abs/1998PhRvC..58.1804A} {58, 1804}

\bibitem[\protect\citeauthoryear{{Alford} \& {Schwenzer}}{{Alford} \&
  {Schwenzer}}{2014a}]{as14_msp}
{Alford} M.~G.,  {Schwenzer} K.,  2014a, \mn@doi [Physical
Review Letters]
  {10.1103/PhysRevLett.113.251102}, \href
  {http://adsabs.harvard.edu/abs/2014PhRvL.113y1102A} {113, 251102}

\bibitem[\protect\citeauthoryear{{Alford} \& {Schwenzer}}{{Alford} \&
  {Schwenzer}}{2014b}]{as14}
{Alford} M.~G.,  {Schwenzer} K.,  2014b, \mn@doi [\apj]
  {10.1088/0004-637X/781/1/26}, \href
  {http://adsabs.harvard.edu/abs/2014ApJ...781...26A} {781, 26}

\bibitem[\protect\citeauthoryear{{Alford} \& {Schwenzer}}{{Alford} \&
  {Schwenzer}}{2015}]{as15_oscMSP}
{Alford} M.~G.,  {Schwenzer} K.,  2015, \mn@doi [\mnras]
  {10.1093/mnras/stu2361}, \href
  {http://adsabs.harvard.edu/abs/2015MNRAS.446.3631A} {446, 3631}

\bibitem[\protect\citeauthoryear{{Alford}, {Mahmoodifar}  \&
  {Schwenzer}}{{Alford} et~al.}{2010}]{ams10}
{Alford} M.~G.,  {Mahmoodifar} S.,   {Schwenzer} K.,  2010,
\mn@doi [Journal of
  Physics G Nuclear Physics] {10.1088/0954-3899/37/12/125202}, \href
  {http://adsabs.harvard.edu/abs/2010JPhG...37l5202A} {37, 125202}

\bibitem[\protect\citeauthoryear{{Alfv\'{e}n} \& {Felthammar}}{{Alfv\'{e}n} \&
  {Felthammar}}{1963}]{AF63}
{Alfv\'{e}n} H.,  {Felthammar} G.~G.,  1963, Cosmical
electrodynamics. Clarendon, Oxford

\bibitem[\protect\citeauthoryear{{Alpar}, {Cheng}, {Ruderman}  \&
  {Shaham}}{{Alpar} et~al.}{1982}]{acrs82}
{Alpar} M.~A.,  {Cheng} A.~F.,  {Ruderman} M.~A.,
{Shaham} J.,  1982, \mn@doi
  [\nat] {10.1038/300728a0}, \href
  {http://adsabs.harvard.edu/abs/1982Natur.300..728A} {300, 728}

\bibitem[\protect\citeauthoryear{{Anderson} \& {Itoh}}{{Anderson} \&
  {Itoh}}{1975}]{ai75}
{Anderson} P.~W.,  {Itoh} N.,  1975, \mn@doi [\nat]
{10.1038/256025a0}, \href
  {http://adsabs.harvard.edu/abs/1975Natur.256...25A} {256, 25}

\bibitem[\protect\citeauthoryear{{Andersson}}{{Andersson}}{1998}]{andersson98}
{Andersson} N.,  1998, \mn@doi [\apj] {10.1086/305919},
\href
  {http://adsabs.harvard.edu/abs/1998ApJ...502..708A} {502, 708}

\bibitem[\protect\citeauthoryear{{Andersson} \& {Comer}}{{Andersson} \&
  {Comer}}{2001}]{ac01}
{Andersson} N.,  {Comer} G.~L.,  2001, \mn@doi [Mon. Not.
R. Astron. Soc.]
  {10.1046/j.1365-8711.2001.04923.x}, \href
  {http://ads.inasan.ru/abs/2001MNRAS.328.1129A} {328, 1129}

\bibitem[\protect\citeauthoryear{{Andersson} \& {Kokkotas}}{{Andersson} \&
  {Kokkotas}}{2001}]{ak01}
{Andersson} N.,  {Kokkotas} K.~D.,  2001, \mn@doi
[International Journal of
  Modern Physics D] {10.1142/S0218271801001062}, \href
  {http://adsabs.harvard.edu/abs/2001IJMPD..10..381A} {10, 381}

\bibitem[\protect\citeauthoryear{{Andersson}, {Jones}  \&
  {Kokkotas}}{{Andersson} et~al.}{2002}]{ajk02}
{Andersson} N.,  {Jones} D.~I.,   {Kokkotas} K.~D.,  2002,
\mn@doi [\mnras]
  {10.1046/j.1365-8711.2002.05837.x}, \href
  {http://adsabs.harvard.edu/abs/2002MNRAS.337.1224A} {337, 1224}

\bibitem[\protect\citeauthoryear{{Andersson}, {Glampedakis}  \&
  {Haskell}}{{Andersson} et~al.}{2009}]{agh09}
{Andersson} N.,  {Glampedakis} K.,   {Haskell} B.,  2009,
\mn@doi [\prd]
  {10.1103/PhysRevD.79.103009}, \href
  {http://adsabs.harvard.edu/abs/2009PhRvD..79j3009A} {79, 103009}

\bibitem[\protect\citeauthoryear{{Andersson}, {Haskell}  \&
  {Comer}}{{Andersson} et~al.}{2010}]{ahc10}
{Andersson} N.,  {Haskell} B.,   {Comer} G.~L.,  2010,
\mn@doi [\prd]
  {10.1103/PhysRevD.82.023007}, \href
  {http://adsabs.harvard.edu/abs/2010PhRvD..82b3007A} {82, 023007}

\bibitem[\protect\citeauthoryear{{Andersson}, {Antonopoulou}, {Espinoza},
  {Haskell}  \& {Ho}}{{Andersson} et~al.}{2017}]{aaehh18}
{Andersson} N.,  {Antonopoulou} D.,  {Espinoza} C.~M.,
{Haskell} B.,   {Ho}
  W.~C.~G.,  2017, preprint, \href
  {http://adsabs.harvard.edu/abs/2017arXiv171105550A} {} (\mn@eprint {arXiv}
  {1711.05550})

\bibitem[\protect\citeauthoryear{{Arras}, {Flanagan}, {Morsink}, {Schenk},
  {Teukolsky}  \& {Wasserman}}{{Arras} et~al.}{2003}]{arras_et_al_03}
{Arras} P.,  {Flanagan} E.~E.,  {Morsink} S.~M.,  {Schenk}
A.~K.,  {Teukolsky}
  S.~A.,   {Wasserman} I.,  2003, \mn@doi [\apj] {10.1086/374657}, \href
  {http://adsabs.harvard.edu/abs/2003ApJ...591.1129A} {591, 1129}

\bibitem[\protect\citeauthoryear{{Asai}, {Lee}  \& {Yoshida}}{{Asai}
  et~al.}{2015}]{aly15}
{Asai} H.,  {Lee} U.,   {Yoshida} S.,  2015, preprint,
\href
  {http://xxx.lanl.gov/abs/1503.04273} {} (\mn@eprint {arXiv} {1503.04273})

\bibitem[\protect\citeauthoryear{{Bhattacharya}, {Heinke}, {Chugunov},
  {Freire}, {Ridolfi}  \& {Bogdanov}}{{Bhattacharya} et~al.}{2017}]{47_Tuc_aa}
{Bhattacharya} S.,  {Heinke} C.~O.,  {Chugunov} A.~I.,
{Freire} P.~C.~C.,
  {Ridolfi} A.,   {Bogdanov} S.,  2017, \mn@doi [\mnras]
  {10.1093/mnras/stx2241}, \href
  {http://adsabs.harvard.edu/abs/2017MNRAS.472.3706B} {472, 3706}

\bibitem[\protect\citeauthoryear{{Bisnovatyi-Kogan} \&
  {Komberg}}{{Bisnovatyi-Kogan} \& {Komberg}}{1976}]{bkk76}
{Bisnovatyi-Kogan} G.~S.,  {Komberg} B.~V.,  1976, Soviet
Astronomy Letters,
  \href {http://adsabs.harvard.edu/abs/1976SvAL....2..130B} {2, 130}

\bibitem[\protect\citeauthoryear{{Bondarescu} \& {Wasserman}}{{Bondarescu} \&
  {Wasserman}}{2013}]{bw13}
{Bondarescu} R.,  {Wasserman} I.,  2013, \mn@doi [\apj]
  {10.1088/0004-637X/778/1/9}, \href
  {http://adsabs.harvard.edu/abs/2013ApJ...778....9B} {778, 9}

\bibitem[\protect\citeauthoryear{{Bondarescu}, {Teukolsky}  \&
  {Wasserman}}{{Bondarescu} et~al.}{2007}]{btw07}
{Bondarescu} R.,  {Teukolsky} S.~A.,   {Wasserman} I.,
2007, \mn@doi [\prd]
  {10.1103/PhysRevD.76.064019}, \href
  {http://adsabs.harvard.edu/abs/2007PhRvD..76f4019B} {76, 064019}

\bibitem[\protect\citeauthoryear{{Bondarescu}, {Teukolsky}  \&
  {Wasserman}}{{Bondarescu} et~al.}{2009}]{btw09}
{Bondarescu} R.,  {Teukolsky} S.~A.,   {Wasserman} I.,
2009, \mn@doi [\prd]
  {10.1103/PhysRevD.79.104003}, \href
  {http://adsabs.harvard.edu/abs/2009PhRvD..79j4003B} {79, 104003}

\bibitem[\protect\citeauthoryear{{Boyer} \& {Lindquist}}{{Boyer} \&
  {Lindquist}}{1966}]{bl66}
{Boyer} R.~H.,  {Lindquist} R.~W.,  1966, \mn@doi [Physics
Letters]
  {10.1016/0031-9163(66)90975-9}, \href
  {http://adsabs.harvard.edu/abs/1966PhL....20..504B} {20, 504}

\bibitem[\protect\citeauthoryear{{Brink}, {Teukolsky}  \& {Wasserman}}{{Brink}
  et~al.}{2004a}]{btw04b}
{Brink} J.,  {Teukolsky} S.~A.,   {Wasserman} I.,  2004a,
\mn@doi [\prd]
  {10.1103/PhysRevD.70.121501}, \href
  {http://adsabs.harvard.edu/abs/2004PhRvD..70l1501B} {70, 121501}

\bibitem[\protect\citeauthoryear{{Brink}, {Teukolsky}  \& {Wasserman}}{{Brink}
  et~al.}{2004b}]{btw04a}
{Brink} J.,  {Teukolsky} S.~A.,   {Wasserman} I.,  2004b,
\mn@doi [\prd]
  {10.1103/PhysRevD.70.124017}, \href
  {http://adsabs.harvard.edu/abs/2004PhRvD..70l4017B} {70, 124017}

\bibitem[\protect\citeauthoryear{{Brown} \& {Cumming}}{{Brown} \&
  {Cumming}}{2009}]{bc09}
{Brown} E.~F.,  {Cumming} A.,  2009, \mn@doi [Astrophys.
J.]
  {10.1088/0004-637X/698/2/1020}, \href
  {http://adsabs.harvard.edu/abs/2009ApJ...698.1020B} {698, 1020}

\bibitem[\protect\citeauthoryear{{Brown}, {Bildsten}  \& {Rutledge}}{{Brown}
  et~al.}{1998}]{bbr98}
{Brown} E.~F.,  {Bildsten} L.,   {Rutledge} R.~E.,  1998,
\mn@doi [\apjl]
  {10.1086/311578}, \href {http://adsabs.harvard.edu/abs/1998ApJ...504L..95B}
  {504, L95}

\bibitem[\protect\citeauthoryear{{Chirenti} \& {Sk{\'a}kala}}{{Chirenti} \&
  {Sk{\'a}kala}}{2013}]{cs13}
{Chirenti} C.,  {Sk{\'a}kala} J.,  2013, \mn@doi [\prd]
  {10.1103/PhysRevD.88.104018}, \href
  {http://adsabs.harvard.edu/abs/2013PhRvD..88j4018C} {88, 104018}

\bibitem[\protect\citeauthoryear{{Chirenti}, {Sk{\'a}kala}  \&
  {Yoshida}}{{Chirenti} et~al.}{2013}]{csy14}
{Chirenti} C.,  {Sk{\'a}kala} J.,   {Yoshida} S.,  2013,
\mn@doi [\prd]
  {10.1103/PhysRevD.87.044043}, \href
  {http://adsabs.harvard.edu/abs/2013PhRvD..87d4043C} {87, 044043}

\bibitem[\protect\citeauthoryear{{Chugunov}}{{Chugunov}}{2015}]{Chugunov15}
{Chugunov} A.~I.,  2015, \mn@doi [\mnras]
{10.1093/mnras/stv1092}, \href
  {http://adsabs.harvard.edu/abs/2015MNRAS.451.2772C} {451, 2772}

\bibitem[\protect\citeauthoryear{{Chugunov}}{{Chugunov}}{2017}]{Chugunov17}
{Chugunov} A.~I.,  2017, \mn@doi [\pasa]
{10.1017/pasa.2017.42}, \href
  {http://adsabs.harvard.edu/abs/2017PASA...34...46C} {34, e046}

\bibitem[\protect\citeauthoryear{{Chugunov}, {Gusakov}  \& {Kantor}}{{Chugunov}
  et~al.}{2014}]{cgk14}
{Chugunov} A.~I.,  {Gusakov} M.~E.,   {Kantor} E.~M.,
2014, \mn@doi [\mnras]
  {10.1093/mnras/stu1772}, \href
  {http://adsabs.harvard.edu/abs/2014MNRAS.445..385C} {445, 385}

\bibitem[\protect\citeauthoryear{Chugunov, Gusakov  \& Kantor}{Chugunov
  et~al.}{2017}]{cgk17}
Chugunov A.~I.,  Gusakov M.~E.,   Kantor E.~M.,  2017,
\mn@doi [\mnras]
  {10.1093/mnras/stx391}, 468, 291

\bibitem[\protect\citeauthoryear{{Chugunov}, {Friedman}, {Lindblom}  \&
    {Rezzolla}}{{Chugunov} et~al.}{2017b}]{cflr17}
{Chugunov} A.~I.,  {Friedman} J.~L.,  {Lindblom} L.,   {Rezzolla} L.,  2017b,
in Journal of Physics Conference Series. p. 012045 (\mn@eprint {arXiv}
{1712.09224}), \mn@doi{10.1088/1742-6596/932/1/012045}

\bibitem[\protect\citeauthoryear{{Dommes} \& {Gusakov}}{{Dommes} \&
  {Gusakov}}{2016}]{dg16}
{Dommes} V.~A.,  {Gusakov} M.~E.,  2016, \mn@doi [MNRAS]
  {10.1093/mnras/stv2408}, \href
  {http://adsabs.harvard.edu/abs/2016MNRAS.455.2852D} {455, 2852}

\bibitem[\protect\citeauthoryear{{Dommes}, {Kantor} \& {Gusakov}}{{Dommes} et~al.}{2018}]{dkg18}
{Dommes} V.~A.,  {Kantor} E.~M., {Gusakov} M.~E.,  2018,
eprint arXiv:1810.08005, \href
{http://adsabs.harvard.edu/abs/2018arXiv181008005D}, accepted for publication in MNRAS



\bibitem[\protect\citeauthoryear{{Flowers} \& {Itoh}}{{Flowers} \&
  {Itoh}}{1979}]{fi79}
{Flowers} E.,  {Itoh} N.,  1979, \mn@doi [\apj]
{10.1086/157145}, \href
  {http://adsabs.harvard.edu/abs/1979ApJ...230..847F} {230, 847}

\bibitem[\protect\citeauthoryear{{Friedman}}{{Friedman}}{1978}]{Friedman78}
{Friedman} J.~L.,  1978, \mn@doi [Communications in
Mathematical Physics]
  {10.1007/BF01202527}, \href
  {http://adsabs.harvard.edu/abs/1978CMaPh..62..247F} {62, 247}

\bibitem[\protect\citeauthoryear{{Friedman} \& {Morsink}}{{Friedman} \&
  {Morsink}}{1998}]{fm98}
{Friedman} J.~L.,  {Morsink} S.~M.,  1998, \mn@doi [\apj]
{10.1086/305920},
  \href {http://adsabs.harvard.edu/abs/1998ApJ...502..714F} {502, 714}

\bibitem[\protect\citeauthoryear{{Friedman} \& {Schutz}}{{Friedman} \&
  {Schutz}}{1978a}]{fs78a}
{Friedman} J.~L.,  {Schutz} B.~F.,  1978a, \mn@doi [\apj]
{10.1086/156098},
  \href {http://adsabs.harvard.edu/abs/1978ApJ...221..937F} {221, 937}

\bibitem[\protect\citeauthoryear{{Friedman} \& {Schutz}}{{Friedman} \&
  {Schutz}}{1978b}]{fs78b}
{Friedman} J.~L.,  {Schutz} B.~F.,  1978b, \mn@doi [\apj]
{10.1086/156143},
  \href {http://adsabs.harvard.edu/abs/1978ApJ...222..281F} {222, 281}

\bibitem[\protect\citeauthoryear{{Friedman} \& {Stergioulas}}{{Friedman} \&
  {Stergioulas}}{2011}]{fs11}
{Friedman} J.~L.,  {Stergioulas} N.,  2011, {Stability of
relativistic stars}. World Scientific Publishing Co, pp
75--100, \mn@doi{10.1142/9789814374774_0007}

\bibitem[\protect\citeauthoryear{Friedman, Lindblom  \& Lockitch}{Friedman
  et~al.}{2016}]{fll16}
Friedman J.~L.,  Lindblom L.,   Lockitch K.~H.,  2016,
\mn@doi [Phys. Rev. D]
  {10.1103/PhysRevD.93.024023}, 93, 024023

\bibitem[\protect\citeauthoryear{{Friedman}, {Lindblom}, {Rezzolla}  \&
  {Chugunov}}{{Friedman} et~al.}{2017}]{flrc17}
{Friedman} J.~L.,  {Lindblom} L.,  {Rezzolla} L.,
{Chugunov} A.~I.,  2017,
  preprint, \href {http://adsabs.harvard.edu/abs/2017arXiv170709419F} {}
  (\mn@eprint {arXiv} {1707.09419})

\bibitem[\protect\citeauthoryear{{Glampedakis} \& {Andersson}}{{Glampedakis} \&
  {Andersson}}{2006a}]{ga06a}
{Glampedakis} K.,  {Andersson} N.,  2006a, \mn@doi [\prd]
  {10.1103/PhysRevD.74.044040}, \href
  {http://adsabs.harvard.edu/abs/2006PhRvD..74d4040G} {74, 044040}

\bibitem[\protect\citeauthoryear{{Glampedakis} \& {Andersson}}{{Glampedakis} \&
  {Andersson}}{2006b}]{ga06b}
{Glampedakis} K.,  {Andersson} N.,  2006b, \mn@doi [\mnras]
  {10.1111/j.1365-2966.2006.10749.x}, \href
  {http://adsabs.harvard.edu/abs/2006MNRAS.371.1311G} {371, 1311}

\bibitem[\protect\citeauthoryear{{Gusakov}, {Yakovlev}  \& {Gnedin}}{{Gusakov}
  et~al.}{2005a}]{gyg05}
{Gusakov} M.~E.,  {Yakovlev} D.~G.,   {Gnedin} O.~Y.,
2005a, \mn@doi [\mnras]
  {10.1111/j.1365-2966.2005.09295.x}, \href
  {http://adsabs.harvard.edu/abs/2005MNRAS.361.1415G} {361, 1415}

\bibitem[\protect\citeauthoryear{{Gusakov}, {Kaminker}, {Yakovlev}  \&
  {Gnedin}}{{Gusakov} et~al.}{2005b}]{gkyg05}
{Gusakov} M.~E.,  {Kaminker} A.~D.,  {Yakovlev} D.~G.,
{Gnedin} O.~Y.,
  2005b, \mn@doi [\mnras] {10.1111/j.1365-2966.2005.09459.x}, \href
  {http://adsabs.harvard.edu/abs/2005MNRAS.363..555G} {363, 555}

\bibitem[\protect\citeauthoryear{{Gusakov}, {Chugunov}  \& {Kantor}}{{Gusakov}
  et~al.}{2014a}]{gck14b}
{Gusakov} M.~E.,  {Chugunov} A.~I.,   {Kantor} E.~M.,
2014a, \mn@doi [\prd]
  {10.1103/PhysRevD.90.063001}, \href
  {http://adsabs.harvard.edu/abs/2014PhRvD..90f3001G} {90, 063001}

\bibitem[\protect\citeauthoryear{{Gusakov}, {Chugunov}  \& {Kantor}}{{Gusakov}
  et~al.}{2014b}]{gck14a}
{Gusakov} M.~E.,  {Chugunov} A.~I.,   {Kantor} E.~M.,
2014b, \mn@doi [\prl]
  {10.1103/PhysRevLett.112.151101}, \href
  {http://adsabs.harvard.edu/abs/2014PhRvL.112o1101G} {112, 151101}

\bibitem[\protect\citeauthoryear{{Gusakov}, {Kantor}  \&
  {Reisenegger}}{{Gusakov} et~al.}{2015}]{gkr15}
{Gusakov} M.~E.,  {Kantor} E.~M.,   {Reisenegger} A.,
2015, \mn@doi [\mnras]
  {10.1093/mnrasl/slv095}, \href
  {http://adsabs.harvard.edu/abs/2015MNRAS.453L..36G} {453, L36}

\bibitem[\protect\citeauthoryear{{Hartle} \& {Sharp}}{{Hartle} \&
  {Sharp}}{1967}]{hs67}
{Hartle} J.~B.,  {Sharp} D.~H.,  1967, \mn@doi [\apj]
{10.1086/149002}, \href
  {http://adsabs.harvard.edu/abs/1967ApJ...147..317H} {147, 317}

\bibitem[\protect\citeauthoryear{{Haskell}}{{Haskell}}{2015}]{haskell15}
{Haskell} B.,  2015, \mn@doi [International Journal of
Modern Physics E]
  {10.1142/S0218301315410074}, \href
  {http://adsabs.harvard.edu/abs/2015IJMPE..2441007H} {24, 1541007}

\bibitem[\protect\citeauthoryear{{Haskell} \& {Patruno}}{{Haskell} \&
  {Patruno}}{2017}]{hp17}
{Haskell} B.,  {Patruno} A.,  2017, \mn@doi [Physical
Review Letters]
  {10.1103/PhysRevLett.119.161103}, \href
  {http://adsabs.harvard.edu/abs/2017PhRvL.119p1103H} {119, 161103}

\bibitem[\protect\citeauthoryear{{Haskell}, {Andersson}  \&
  {Passamonti}}{{Haskell} et~al.}{2009}]{hap09}
{Haskell} B.,  {Andersson} N.,   {Passamonti} A.,  2009,
\mn@doi [\mnras]
  {10.1111/j.1365-2966.2009.14963.x}, \href
  {http://adsabs.harvard.edu/abs/2009MNRAS.397.1464H} {397, 1464}

\bibitem[\protect\citeauthoryear{{Haskell}, {Glampedakis}  \&
  {Andersson}}{{Haskell} et~al.}{2014}]{hga14}
{Haskell} B.,  {Glampedakis} K.,   {Andersson} N.,  2014,
\mn@doi [\mnras]
  {10.1093/mnras/stu535}, \href
  {http://adsabs.harvard.edu/abs/2014MNRAS.441.1662H} {441, 1662}

\bibitem[\protect\citeauthoryear{{Heiselberg} \& {Hjorth-Jensen}}{{Heiselberg}
  \& {Hjorth-Jensen}}{1999}]{hh99}
{Heiselberg} H.,  {Hjorth-Jensen} M.,  1999, \mn@doi
[\apjl] {10.1086/312321},
  \href {http://adsabs.harvard.edu/abs/1999ApJ...525L..45H} {525, L45}

\bibitem[\protect\citeauthoryear{{Ho} \& {Lai}}{{Ho} \& {Lai}}{2000}]{hl00}
{Ho} W.~C.~G.,  {Lai} D.,  2000, \mn@doi [\apj]
{10.1086/317085}, \href
  {http://adsabs.harvard.edu/abs/2000ApJ...543..386H} {543, 386}

\bibitem[\protect\citeauthoryear{Jackson}{Jackson}{1999}]{JacksonCED}
Jackson J.~D.,  1999, Classical electrodynamics, 3rd ed.
edn. Wiley, New York, {NY}, \url
{http://cdsweb.cern.ch/record/490457}

\bibitem[\protect\citeauthoryear{{Jasiulek} \& {Chirenti}}{{Jasiulek} \&
  {Chirenti}}{2017}]{jc17}
{Jasiulek} M.,  {Chirenti} C.,  2017, \mn@doi [\prd]
  {10.1103/PhysRevD.95.064060}, \href
  {http://adsabs.harvard.edu/abs/2017PhRvD..95f4060J} {95, 064060}

\bibitem[\protect\citeauthoryear{{Jones}}{{Jones}}{2001}]{Jones01_comment}
{Jones} P.~B.,  2001, \mn@doi [Physical Review Letters]
  {10.1103/PhysRevLett.86.1384}, \href
  {http://adsabs.harvard.edu/abs/2001PhRvL..86.1384J} {86, 1384}

\bibitem[\protect\citeauthoryear{{Kantor} \& {Gusakov}}{{Kantor} \&
  {Gusakov}}{2014}]{kg14}
{Kantor} E.~M.,  {Gusakov} M.~E.,  2014, \mn@doi [MNRAS]
  {10.1093/mnrasl/slu061}, \href
  {http://adsabs.harvard.edu/abs/2014MNRAS.442L..90K} {442, L90}

\bibitem[\protect\citeauthoryear{{Kantor} \& {Gusakov}}{{Kantor} \&
  {Gusakov}}{2017}]{kg17}
{Kantor} E.~M.,  {Gusakov} M.~E.,  2017, \mn@doi [\mnras]
  {10.1093/mnras/stx1075}, \href
  {http://adsabs.harvard.edu/abs/2017MNRAS.469.3928K} {469, 3928}

\bibitem[\protect\citeauthoryear{{Kantor}, {Gusakov}  \& {Chugunov}}{{Kantor}
  et~al.}{2016}]{kgc16}
{Kantor} E.~M.,  {Gusakov} M.~E.,   {Chugunov} A.~I.,
2016, \mn@doi [\mnras]
  {10.1093/mnras/stv2352}, \href
  {http://adsabs.harvard.edu/abs/2016MNRAS.455..739K} {455, 739}

\bibitem[\protect\citeauthoryear{{Kastaun}}{{Kastaun}}{2011}]{Kastaun11}
{Kastaun} W.,  2011, \mn@doi [\prd]
{10.1103/PhysRevD.84.124036}, \href
  {http://adsabs.harvard.edu/abs/2011PhRvD..84l4036K} {84, 124036}

\bibitem[\protect\citeauthoryear{Kokkotas \& Schwenzer}{Kokkotas \&
  Schwenzer}{2016}]{ks16}
Kokkotas K.~D.,  Schwenzer K.,  2016, \mn@doi [The European
Physical Journal A]
  {10.1140/epja/i2016-16038-9}, 52, 38

\bibitem[\protect\citeauthoryear{{Kolomeitsev} \& {Voskresensky}}{{Kolomeitsev}
  \& {Voskresensky}}{2015}]{kv15}
{Kolomeitsev} E.~E.,  {Voskresensky} D.~N.,  2015, \mn@doi
[\prc]
  {10.1103/PhysRevC.91.025805}, \href
  {http://adsabs.harvard.edu/abs/2015PhRvC..91b5805K} {91, 025805}

\bibitem[\protect\citeauthoryear{{Kr{\"u}ger}, {Gaertig}  \&
  {Kokkotas}}{{Kr{\"u}ger} et~al.}{2010}]{kgk10}
{Kr{\"u}ger} C.,  {Gaertig} E.,   {Kokkotas} K.~D.,  2010,
\mn@doi [\prd]
  {10.1103/PhysRevD.81.084019}, \href
  {http://adsabs.harvard.edu/abs/2010PhRvD..81h4019K} {81, 084019}

\bibitem[\protect\citeauthoryear{Landau \& Lifshitz}{Landau \&
  Lifshitz}{1987}]{ll87}
Landau L.~D.,  Lifshitz E.,  1987, Fluid mechanics. Course
of theoretical
  physics.
Pergamon Press, Oxford

\bibitem[\protect\citeauthoryear{{Lee}}{{Lee}}{2005}]{Lee05}
{Lee} U.,  2005, \mn@doi [\mnras]
{10.1111/j.1365-2966.2004.08614.x}, \href
  {http://adsabs.harvard.edu/abs/2005MNRAS.357...97L} {357, 97}

\bibitem[\protect\citeauthoryear{{Lee} \& {Yoshida}}{{Lee} \&
  {Yoshida}}{2003}]{ly03}
{Lee} U.,  {Yoshida} S.,  2003, \mn@doi [\apj]
{10.1086/367617}, \href
  {http://adsabs.harvard.edu/abs/2003ApJ...586..403L} {586, 403}

\bibitem[\protect\citeauthoryear{{Levin}}{{Levin}}{1999}]{levin99}
{Levin} Y.,  1999, \mn@doi [\apj] {10.1086/307196}, \href
  {http://adsabs.harvard.edu/abs/1999ApJ...517..328L} {517, 328}

\bibitem[\protect\citeauthoryear{{Levin} \& {Ushomirsky}}{{Levin} \&
  {Ushomirsky}}{2001a}]{lu01b}
{Levin} Y.,  {Ushomirsky} G.,  2001a, \mn@doi [\mnras]
  {10.1046/j.1365-8711.2001.04075.x}, \href
  {http://adsabs.harvard.edu/abs/2001MNRAS.322..515L} {322, 515}

\bibitem[\protect\citeauthoryear{{Levin} \& {Ushomirsky}}{{Levin} \&
  {Ushomirsky}}{2001b}]{lu01}
{Levin} Y.,  {Ushomirsky} G.,  2001b, \mn@doi [\mnras]
  {10.1046/j.1365-8711.2001.04323.x}, \href
  {http://adsabs.harvard.edu/abs/2001MNRAS.324..917L} {324, 917}

\bibitem[\protect\citeauthoryear{{Lindblom} \& {Mendell}}{{Lindblom} \&
  {Mendell}}{2000}]{lm00}
{Lindblom} L.,  {Mendell} G.,  2000, \mn@doi [\prd]
  {10.1103/PhysRevD.61.104003}, \href
  {http://adsabs.harvard.edu/abs/2000PhRvD..61j4003L} {61, 104003}

\bibitem[\protect\citeauthoryear{{Lindblom} \& {Owen}}{{Lindblom} \&
  {Owen}}{2002}]{lo02}
{Lindblom} L.,  {Owen} B.~J.,  2002, \mn@doi [\prd]
  {10.1103/PhysRevD.65.063006}, \href
  {http://adsabs.harvard.edu/abs/2002PhRvD..65f3006L} {65, 063006}

\bibitem[\protect\citeauthoryear{{Lindblom}, {Owen}  \& {Morsink}}{{Lindblom}
  et~al.}{1998}]{lom98}
{Lindblom} L.,  {Owen} B.~J.,   {Morsink} S.~M.,  1998,
\mn@doi [\prl]
  {10.1103/PhysRevLett.80.4843}, \href
  {http://adsabs.harvard.edu/abs/1998PhRvL..80.4843L} {80, 4843}

\bibitem[\protect\citeauthoryear{{Lindblom}, {Mendell}  \& {Owen}}{{Lindblom}
  et~al.}{1999}]{lmo99}
{Lindblom} L.,  {Mendell} G.,   {Owen} B.~J.,  1999,
\mn@doi [\prd]
  {10.1103/PhysRevD.60.064006}, \href
  {http://adsabs.harvard.edu/abs/1999PhRvD..60f4006L} {60, 064006}

\bibitem[\protect\citeauthoryear{{Lockitch}, {Friedman}  \&
  {Andersson}}{{Lockitch} et~al.}{2003}]{lfa03}
{Lockitch} K.~H.,  {Friedman} J.~L.,   {Andersson} N.,
2003, \mn@doi [\prd]
  {10.1103/PhysRevD.68.124010}, \href
  {http://adsabs.harvard.edu/abs/2003PhRvD..68l4010L} {68, 124010}

\bibitem[\protect\citeauthoryear{{Longuet-Higgins}}{{Longuet-Higgins}}{1953}]{LH53}
{Longuet-Higgins} M.~S.,  1953, Phil. Trans. R. Soc. Long.
A, 245, 535

\bibitem[\protect\citeauthoryear{{Mahmoodifar} \& {Strohmayer}}{{Mahmoodifar}
  \& {Strohmayer}}{2013}]{ms13}
{Mahmoodifar} S.,  {Strohmayer} T.,  2013, \mn@doi [\apj]
  {10.1088/0004-637X/773/2/140}, \href
  {http://adsabs.harvard.edu/abs/2013ApJ...773..140M} {773, 140}

\bibitem[\protect\citeauthoryear{{McDermott}, {Savedoff}, {van Horn}, {Zweibel}
   \& {Hansen}}{{McDermott} et~al.}{1984}]{McDermott_etal84_OscEmEmis}
{McDermott} P.~N.,  {Savedoff} M.~P.,  {van Horn} H.~M.,
{Zweibel} E.~G.,
  {Hansen} C.~J.,  1984, \mn@doi [\apj] {10.1086/162152}, \href
  {http://adsabs.harvard.edu/abs/1984ApJ...281..746M} {281, 746}

\bibitem[\protect\citeauthoryear{{Meisel}, {Deibel}, {Keek}, {Shternin}  \&
  {Elfritz}}{{Meisel} et~al.}{2018}]{mdkse18}
{Meisel} Z.,  {Deibel} A.,  {Keek} L.,  {Shternin} P.,
{Elfritz} J.,  2018,
  preprint, \href {http://adsabs.harvard.edu/abs/2018arXiv180701150M} {}
  (\mn@eprint {arXiv} {1807.01150})

\bibitem[\protect\citeauthoryear{{Misner}, {Thorne}  \& {Wheeler}}{{Misner}
  et~al.}{1973}]{MTW}
{Misner} C.~W.,  {Thorne} K.~S.,   {Wheeler} J.~A.,  1973,
{Gravitation}. W.H.~Freeman and Co., San Francisco

\bibitem[\protect\citeauthoryear{{Nayyar} \& {Owen}}{{Nayyar} \&
  {Owen}}{2006}]{no06}
{Nayyar} M.,  {Owen} B.~J.,  2006, \mn@doi [\prd]
{10.1103/PhysRevD.73.084001},
  \href {http://adsabs.harvard.edu/abs/2006PhRvD..73h4001N} {73, 084001}

\bibitem[\protect\citeauthoryear{{Ofengeim}, {Fortin}, {Haensel}, {Yakovlev}
  \& {Zdunik}}{{Ofengeim} et~al.}{2017}]{Ofengeim_etal17}
{Ofengeim} D.~D.,  {Fortin} M.,  {Haensel} P.,  {Yakovlev}
D.~G.,   {Zdunik}
  J.~L.,  2017, \mn@doi [\prd] {10.1103/PhysRevD.96.043002}, \href
  {http://adsabs.harvard.edu/abs/2017PhRvD..96d3002O} {96, 043002}

\bibitem[\protect\citeauthoryear{{Ootes}, {Page}, {Wijnands}  \&
  {Degenaar}}{{Ootes} et~al.}{2016}]{Ootes_etal16}
{Ootes} L.~S.,  {Page} D.,  {Wijnands} R.,   {Degenaar} N.,
2016, \mn@doi
  [\mnras] {10.1093/mnras/stw1799}, \href
  {http://adsabs.harvard.edu/abs/2016MNRAS.461.4400O} {461, 4400}

\bibitem[\protect\citeauthoryear{{Owen}}{{Owen}}{2010}]{owen10}
{Owen} B.~J.,  2010, \mn@doi [\prd]
{10.1103/PhysRevD.82.104002}, \href
  {http://adsabs.harvard.edu/abs/2010PhRvD..82j4002O} {82, 104002}

\bibitem[\protect\citeauthoryear{{Owen}, {Lindblom}, {Cutler}, {Schutz},
  {Vecchio}  \& {Andersson}}{{Owen} et~al.}{1998}]{olcsva98}
{Owen} B.~J.,  {Lindblom} L.,  {Cutler} C.,  {Schutz}
B.~F.,  {Vecchio} A.,
  {Andersson} N.,  1998, \mn@doi [\prd] {10.1103/PhysRevD.58.084020}, \href
  {http://adsabs.harvard.edu/abs/1998PhRvD..58h4020O} {58, 084020}

\bibitem[\protect\citeauthoryear{{Page}, {Lattimer}, {Prakash}  \&
  {Steiner}}{{Page} et~al.}{2004}]{plps04}
{Page} D.,  {Lattimer} J.~M.,  {Prakash} M.,   {Steiner}
A.~W.,  2004, \mn@doi
  [\apjs] {10.1086/424844}, \href
  {http://adsabs.harvard.edu/abs/2004ApJS..155..623P} {155, 623}

\bibitem[\protect\citeauthoryear{{Papaloizou} \& {Pringle}}{{Papaloizou} \&
  {Pringle}}{1978}]{pp78}
{Papaloizou} J.,  {Pringle} J.~E.,  1978, \mnras, \href
  {http://adsabs.harvard.edu/abs/1978MNRAS.182..423P} {182, 423}

\bibitem[\protect\citeauthoryear{{Parikh}, {Wijnands}, {Degenaar}, {Ootes}  \&
  {Page}}{{Parikh} et~al.}{2018}]{Parikh_etal18_crustcool}
{Parikh} A.~S.,  {Wijnands} R.,  {Degenaar} N.,  {Ootes}
L.,   {Page} D.,
  2018, \mn@doi [\mnras] {10.1093/mnras/sty416}, \href
  {http://adsabs.harvard.edu/abs/2018MNRAS.476.2230P} {476, 2230}

\bibitem[\protect\citeauthoryear{{Patruno} \& {Watts}}{{Patruno} \& {Watts}}{2012}]{pw12}
Patruno A., Watts A. L., 2012, preprint (arXiv:1206.2727)

\bibitem[\protect\citeauthoryear{{Price} \& {Thorne}}{{Price} \&
  {Thorne}}{1969}]{pt69}
{Price} R.,  {Thorne} K.~S.,  1969, \mn@doi [\apj]
{10.1086/149857}, \href
  {http://adsabs.harvard.edu/abs/1969ApJ...155..163P} {155, 163}

\bibitem[\protect\citeauthoryear{{Reisenegger} \& {Bona{\v
  c}i{\'c}}}{{Reisenegger} \& {Bona{\v c}i{\'c}}}{2003a}]{rb03_bulk}
{Reisenegger} A.,  {Bona{\v c}i{\'c}} A.,  2003a, in
{Cusumano} G.,  {Massaro}
  E.,   {Mineo} T.,  eds, Pulsars, AXPs and SGRs Observed with BeppoSAX and
  Other Observatories. pp 231--236 (\mn@eprint {} {astro-ph/0303454})

\bibitem[\protect\citeauthoryear{{Reisenegger} \& {Bona{\v
  c}i{\'c}}}{{Reisenegger} \& {Bona{\v c}i{\'c}}}{2003b}]{rb03}
{Reisenegger} A.,  {Bona{\v c}i{\'c}} A.,  2003b, \mn@doi
[\prl]
  {10.1103/PhysRevLett.91.201103}, \href
  {http://adsabs.harvard.edu/abs/2003PhRvL..91t1103R} {91, 201103}

\bibitem[\protect\citeauthoryear{{Rezzolla}, {Lamb}  \& {Shapiro}}{{Rezzolla}
  et~al.}{2000}]{Rezzolla_etal00}
{Rezzolla} L.,  {Lamb} F.~K.,   {Shapiro} S.~L.,  2000,
\mn@doi [\apjl]
  {10.1086/312539}, \href {http://adsabs.harvard.edu/abs/2000ApJ...531L.139R}
  {531, L139}

\bibitem[\protect\citeauthoryear{{Rezzolla}, {Lamb}, {Markovi{\'c}}  \&
  {Shapiro}}{{Rezzolla} et~al.}{2001}]{Rezzolla_etal01a}
{Rezzolla} L.,  {Lamb} F.~K.,  {Markovi{\'c}} D.,
{Shapiro} S.~L.,  2001,
  \mn@doi [\prd] {10.1103/PhysRevD.64.104013}, \href
  {http://adsabs.harvard.edu/abs/2001PhRvD..64j4013R} {64, 104013}

\bibitem[\protect\citeauthoryear{{Rieutord}}{{Rieutord}}{2001}]{rieutord01}
{Rieutord} M.,  2001, \mn@doi [\apj] {10.1086/319705},
\href
  {http://adsabs.harvard.edu/abs/2001ApJ...550..443R} {550, 443}

\bibitem[\protect\citeauthoryear{{Ruoff} \& {Kokkotas}}{{Ruoff} \&
  {Kokkotas}}{2002}]{rk02}
{Ruoff} J.,  {Kokkotas} K.~D.,  2002, \mn@doi [\mnras]
  {10.1046/j.1365-8711.2002.05169.x}, \href
  {http://adsabs.harvard.edu/abs/2002MNRAS.330.1027R} {330, 1027}

\bibitem[\protect\citeauthoryear{Rutledge, Bildsten, Brown, Pavlov, Zavlin  \&
  Ushomirsky}{Rutledge et~al.}{2002}]{Ruthledge_etal02_KS}
Rutledge R.~E.,  Bildsten L.,  Brown E.~F.,  Pavlov G.~G.,
Zavlin V.~E.,
  Ushomirsky G.,  2002, The Astrophysical Journal, 580, 413

\bibitem[\protect\citeauthoryear{{S{\'a}}}{{S{\'a}}}{2004}]{Sa04}
{S{\'a}} P.~M.,  2004, \mn@doi [\prd]
{10.1103/PhysRevD.69.084001}, \href
  {http://adsabs.harvard.edu/abs/2004PhRvD..69h4001S} {69, 084001}

\bibitem[\protect\citeauthoryear{{Schmitt} \& {Shternin}}{{Schmitt} \&
  {Shternin}}{2017}]{ss17}
{Schmitt} A.,  {Shternin} P.,  2017, preprint, \href
  {http://adsabs.harvard.edu/abs/2017arXiv171106520S} {} (\mn@eprint {arXiv}
  {1711.06520})

\bibitem[\protect\citeauthoryear{{Schwenzer}, {Boztepe}, {G{\"u}ver}  \&
  {Vurgun}}{{Schwenzer} et~al.}{2017}]{Schwenzer_etal_Xray}
{Schwenzer} K.,  {Boztepe} T.,  {G{\"u}ver} T.,   {Vurgun}
E.,  2017, \mn@doi
  [\mnras] {10.1093/mnras/stw3201}, \href
  {http://adsabs.harvard.edu/abs/2017MNRAS.466.2560S} {466, 2560}

%\bibitem[\protect\citeauthoryear{{Shternin}}{{Shternin}}{2018}]{Shternin18}
%{Shternin} P.~S.,  2018, \mn@doi [\prd]
%{10.1103/PhysRevD.98.063015}, \href
%{http://adsabs.harvard.edu/abs/2018PhRvD..98f3015S} {98, 063015}


\bibitem[\protect\citeauthoryear{{Shternin} \& {Yakovlev}}{{Shternin} \&
  {Yakovlev}}{2008}]{sy08}
{Shternin} P.~S.,  {Yakovlev} D.~G.,  2008, \mn@doi [\prd]
  {10.1103/PhysRevD.78.063006}, \href
  {http://adsabs.harvard.edu/abs/2008PhRvD..78f3006S} {78, 063006}

\bibitem[\protect\citeauthoryear{{Shternin}, {Yakovlev}, {Haensel}  \&
  {Potekhin}}{{Shternin} et~al.}{2007}]{syhp07}
{Shternin} P.~S.,  {Yakovlev} D.~G.,  {Haensel} P.,
{Potekhin} A.~Y.,  2007,
  \mn@doi [Mon. Not. R. Astron. Soc.] {10.1111/j.1745-3933.2007.00386.x}, \href
  {http://adsabs.harvard.edu/abs/2007MNRAS.382L..43S} {382, L43}

\bibitem[\protect\citeauthoryear{{Shternin}, {Baldo}  \& {Schulze}}{{Shternin}
  et~al.}{2017}]{sbs17}
{Shternin} P.~S.,  {Baldo} M.,   {Schulze} H.-J.,  2017, in
Journal of Physics
  Conference Series. p. 012042 (\mn@eprint {arXiv} {1711.00371}),
  \mn@doi{10.1088/1742-6596/932/1/012042}

\bibitem[\protect\citeauthoryear{{Spruit}}{{Spruit}}{1999}]{Spruit99_DifRot}
{Spruit} H.~C.,  1999, \aap, \href
  {http://adsabs.harvard.edu/abs/1999A%26A...349..189S} {349, 189}

\bibitem[\protect\citeauthoryear{{Stergioulas}}{{Stergioulas}}{2003}]{Stergioulas03}
{Stergioulas} N.,  2003, \mn@doi [Living Reviews in
Relativity]
  {10.12942/lrr-2003-3}, \href
  {http://adsabs.harvard.edu/abs/2003LRR.....6....3S} {6, 3}

\bibitem[\protect\citeauthoryear{Stokes}{Stokes}{1847}]{Stokes1847}
Stokes G.~G.,  1847, Transactions of the Cambridge
Philosophical Society, 8,
  441

\bibitem[\protect\citeauthoryear{{The LIGO Scientific Collaboration}
  et~al.,}{{The LIGO Scientific Collaboration} et~al.}{2018}]{LIGO18_AllSky}
{The LIGO Scientific Collaboration} et~al., 2018, preprint,
\href
  {http://adsabs.harvard.edu/abs/2018arXiv180205241T} {} (\mn@eprint {arXiv}
  {1802.05241})

\bibitem[\protect\citeauthoryear{{Thorne}}{{Thorne}}{1980}]{Thorne80}
{Thorne} K.~S.,  1980, \mn@doi [Reviews of Modern Physics]
  {10.1103/RevModPhys.52.299}, \href
  {http://adsabs.harvard.edu/abs/1980RvMP...52..299T} {52, 299}

\bibitem[\protect\citeauthoryear{Varshalovich, Moskalev  \&
  Khersonskii}{Varshalovich et~al.}{1988}]{vms88}
Varshalovich D.,  Moskalev A.,   Khersonskii V.,  1988,
Quantum Theory of
  Angular Momentum.
World Scientific Pub., Singapore, \url
  {https://books.google.ru/books?id=nXcGCwAAQBAJ}

\bibitem[\protect\citeauthoryear{Wijnands, Degenaar  \& Page}{Wijnands
  et~al.}{2017}]{wdp17}
Wijnands R.,  Degenaar N.,   Page D.,  2017, \mn@doi
[Journal of Astrophysics
  and Astronomy] {10.1007/s12036-017-9466-5}, 38, 49

\bibitem[\protect\citeauthoryear{{Yakovlev} \& {Pethick}}{{Yakovlev} \&
  {Pethick}}{2004}]{yp04}
{Yakovlev} D.~G.,  {Pethick} C.~J.,  2004, \mn@doi [\araa]
  {10.1146/annurev.astro.42.053102.134013}, \href
  {http://adsabs.harvard.edu/abs/2004ARA%26A..42..169Y} {42, 169}

\bibitem[\protect\citeauthoryear{{Yoshida} \& {Lee}}{{Yoshida} \&
  {Lee}}{2003a}]{yl03b}
{Yoshida} S.,  {Lee} U.,  2003a, \mn@doi [\prd]
{10.1103/PhysRevD.67.124019},
  \href {http://adsabs.harvard.edu/abs/2003PhRvD..67l4019Y} {67, 124019}

\bibitem[\protect\citeauthoryear{{Yoshida} \& {Lee}}{{Yoshida} \&
  {Lee}}{2003b}]{yl03a}
{Yoshida} S.,  {Lee} U.,  2003b, \mn@doi [\mnras]
  {10.1046/j.1365-8711.2003.06816.x}, \href
  {http://adsabs.harvard.edu/abs/2003MNRAS.344..207Y} {344, 207}

\makeatother
\end{thebibliography}
\end{document}